\documentclass[prb,superscriptaddress,twocolumn,amsmath,amssymb]{revtex4}
\usepackage{graphicx}
\usepackage{bm}
\usepackage{physics}
\usepackage{MnSymbol}
\usepackage{color}
\usepackage{enumerate}

\begin{document}
\newcommand{\normord}[1]{\vcentcolon\mathrel{#1}\vcentcolon}
\providecommand{\vcentcolon}{\mathrel{\mathop{:}}}
\title{Effective theory of lattice electrons  strongly coupled to quantum electromagnetic fields}
\author{Jiajun Li}
\affiliation{Chair of Theoretical Solid-State Physics, University of Erlangen-Nuremberg, 91054 Erlangen}
\affiliation{Laboratory for Theoretical and Computational Physics, Paul Scherrer Institute, 5232 PSI Villigen, Switzerland}
\author{Lukas Schamri\ss}
\affiliation{Chair of Theoretical Solid-State Physics, University of Erlangen-Nuremberg, 91054 Erlangen}
\author{Martin Eckstein}
\affiliation{Chair of Theoretical Solid-State Physics, University of Erlangen-Nuremberg, 91054 Erlangen}
\begin{abstract}
Recent experiments have revealed the tantalizing possibility of fabricating lattice electronic systems strongly coupled to quantum fluctuations of electromagnetic fields, e.g., by means of geometry confinement from a cavity or artificial gauge fields in quantum simulators. In this work, we develop a high-frequency expansion to construct the effective models for lattice electrons strongly coupled to a continuum of off-resonant photon modes with arbitrary dispersion. The theory is nonperturbative in the light-matter coupling strength, and is therefore particularly suitable for the ultrastrong light-matter coupling regime. Using the effective models, we demonstrate how the dispersion and topology of the electronic energy bands can be tuned by the cavity. In particular, quasi-one-dimensional physics can emerge in a two-dimensional square lattice due to a spatially anisotropic band renormalization, and a topologically nontrivial anomalous quantum Hall state can be induced in a honeycomb lattice when the cavity setup breaks time-reversal symmetry. We also demonstrate that the photon-mediated interaction induces an unconventional superconducting paired phase distinct from the pair-density-wave state discussed in models with truncated light-matter coupling. Finally, we study a realistic setup of a Fabry-P\'{e}rot cavity. Our work provides a systematic framework to explore the emergent phenomena due to strong light-matter coupling and points out new directions of engineering orders and topological states in solids. 
\end{abstract}
\maketitle

\section{Introduction}
The electromagnetic field, as a prototypical example of a $U(1)$ gauge field, gives rise to the fundamental interaction between charged particles in condensed matter and atomic physics. Ultrafast lasers have been used as a versatile tool to engineer the transient properties of materials \cite{basov2017, torre2021}, since the strong and coherent electromagnetic field can interact with essentially all degrees of freedom in solids, including charge \cite{stojchevska2014,ligges2018}, orbital \cite{ichikawa2011}, spin \cite{mikhaylovskiy2015} and lattice \cite{forst2011,singla2015,esposito2017,kozina2019}. Quantum fluctuations of the electromagnetic field are often negligible in these scenarios, due to the weak electron-photon coupling in vacuum controlled by fine structure constant $\alpha$. However, recent experimental advances have offered the prospect of strongly enhanced electron-photon coupling \cite{forn-diaz2019}. Possible implementations include condensed matters inside subwavelength cavities \cite{scalari2012,geiser2012,maissen2014}, with possible cavity-controlled superconductivity reported in a recent experiment \cite{thomas2019}, and artificial dynamical gauge fields in scalable quantum simulators \cite{gorg2019,mil2020,yang2020}. In the context of solid-state physics, strong coupling of lattice fermions and gauge bosons can also emerge in doped Mott insulators \cite{lee2006, lee2014}. 

At strong coupling, one can expect hybrid light-matter phases similar to quantum optics, where the strong hybridization of atomic states and cavity photons leads to intriguing phenomena such as superradiance \cite{dicke1954,hepp1973, mivehvar2021}. This is an exotic quantum optical phenomenon which is completely absent in the classical-light driving regimes. Moreover, the strong light-matter coupling holds the promise of implementing \emph{equilibrium analogs} of laser control of material properties, such as transient band renormalization \cite{schmitt2008} and nonthermal topological phases \cite{wang2013,mciver2020}. The hybrid light-matter phases therefore provide a potential pathway to realize 
 Floquet engineered states or light-induced hidden phases as a stable thermal state \cite{torre2021,schlawin2021}, minimizing the detrimental laser heating in a driven system. First theoretical studies have unveiled the tantalizing possibility of cavity-enhanced ferroelectrics \cite{ashida2020} and excitonic condensates \cite{lenk2020,mazza2019}, extending the debate on equilibrium superradiant transition to condensed matter systems \cite{mazza2019,andolina2019,andolina2020,guerci2020}. The hybrid light-matter phases can also feature distinct magnetic order \cite{kiffner2019,sentef2020,chiocchetta2020,rohn2020} and superconducting pairing mechanisms \cite{sentef2018,curtis2019,schlawin2019,schlawin2019b,gao2020,chakraborty2020,li2020prl}. 

An attractive possibility is to engineer the electronic band structure using quantum light, in analogy to laser-induced band engineering which is currently under intense scrutiny \cite{oka2009,kitagawa2011,lindner2011, oka2019, rudner2020, wang2013,mciver2020}. Recent theoretical works, e.g., predict a mass renormalization for the free electrons gas in a Fabry-P\'{e}rot cavity \cite{rokaj2020}. For lattice systems, however, a particular difficulty in theory is that the gauge-invariant light-matter coupling \footnote{In the Hamiltonian formalism, the electromagnetic fields are always quantized in a fixed gauge. However, the light-matter theory can be recast into a gauge-invariant path integral (Lagrangian) formulation only when the full nonlinearity of the Peierls phase is retained.} involves the highly nonlinear Peierls phase factor \cite{boykin2001,golez2019,li2020prb,schueler2021}, except for intrinsically nondispersive bands \cite{schachenmayer2015,hagenmueller2017}. Moreover, the continuous band structure is essentially a property of extended systems, and a continuum of photon modes, corresponding to a spatially varying $U(1)$ gauge field, must be taken into account in the thermodynamic limit \cite{rokaj2020}. The coupling to a continuum of modes also holds the promise of scaling up the cavity-induced effects in small-size systems, which become no longer limited by the coupling strength of a single photon mode \cite{kockum2019}, as well as overcoming the no-go theorem of equilibrium superradiance \cite{andolina2020, guerci2020}. The combination of a continuous quantum gauge field with a highly nonlinear gauge-invariant coupling places a formidable challenge for theoretical studies. 

\begin{figure}
\includegraphics[scale=1.7]{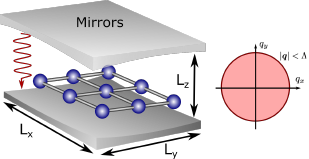}
\caption{Cavity engineering of solid-state systems. The left panel shows a layer of the material sample placed inside a cavity. The dimensions of the cavity are labelled by $L_x,L_y,L_z$, where $L_x,L_y\to\infty$ is assumed. The right panel schematically shows the continuum of modes coupled to the lattice electrons of different transverse momenta. A UV-cutoff $\Lambda$ in $\bm q$ is necessary for regularizing ultraviolet divergence in quantum electrodynamics.}
\label{fig:cavity}
\end{figure}

In this article, we formulate a framework to describe lattice electrons coupled to a continuum of boson modes using a high-frequency expansion \cite{bukov2015,mikami2016}, in analogy to the Floquet formalism for classical-laser driving. Since our purpose is to provide a general framework 
for systems with strong light-matter coupling, we assume an arbitrary cavity geometry where fields are confined in $z$-direction, but extended in $x$ and $y$ direction. The simplest implementation is to place a material between two metallic mirrors, as in a Fabry-P\'{e}rot cavity, but one can also imagine more complicated structures, such as scalable nanoplasmonic cavities which can acquire a higher compression of the fields in $z$ direction. In the thermodynamic limit $L_x,L_y\to\infty$, the electronic system is coupled to a continuum of modes labelled by the in-plane-momentum $(q_x, q_y)$, and a discrete mode index due to the confinement in the $z$--direction and the polarization. For example, in a Fabry-P\'{e}rot cavity the transverse electric field must vanish at the cavity boundaries, leading to quantization of $z$-momentum $q_z=n_zq_c$, with lowest momentum $q_c=\pi/L_z$, and a corresponding cavity frequency $\omega_c=q_cv$, given the speed of light $v$. The mode frequency is then given by $\omega_{\bm q}=\sqrt{n_z^2\omega_c^2+v^2(q_x^2+q_y^2)}$. Typically, the cavity is able to confine photon modes below a certain energy scale, such as the plasma frequency of the metallic plates. 
 
It is worth noting that placing the matter inside the cavity does not suddenly create the light-matter coupling, but it rather changes the mode structure from the free-space situation to the confined one. In the theoretical interpretation of the cavity-induced effect on matter, we can therefore distinguish two situations: In a subwavelength cavity, the coupling to a few particular modes can be significantly enhanced compared to the free-space situation. This situation can thus be described by taking the uncoupled Hamiltonian $H_0$ to be an empirical free-space matter Hamiltonian, with vacuum photon dressing already included, coupled to only the prominently enhanced modes \cite{schlawin2019}. The cavity-induced effect on the solid is accounted for as the dressing due to these particular strongly coupled mode(s). We will refer to this as Situation I in the following. In contrast, Situation II refers to the case when the cavity modifies the mode structure, but does not particularly enhance the coupling strength to individual modes, as for a conventional Fabry-P\'{e}rot cavity. The dressing of material parameters due to all modes in a Fabry-P\'{e}rot cavity (possibly below some cutoff set by the  plasma frequency of the mirrors) would remain nonzero in the free space limit ($L_z\to \infty$, $\omega_c\to 0$), and the description of the light-matter states in this situation requires calculating the difference of the photon-dressing in vacuum and in the cavity. In this situation one would take the uncoupled Hamiltonian $H_0$ to be the bare electronic Hamiltonian $H_{\rm bare}$, and include the coupling to all photon modes inside a cavity. If one integrates out the photon degree of freedom, the renormalized Hamiltonian $H_{\rm ren}(\omega_c)$ then depends on the cavity frequency $\omega_c$. The free-space limit is approached by taking $\omega_c\to0$, i.e., $H_{\rm free}=\lim_{\omega_c\to0}H_{\rm ren}(\omega_c)$. The role of the cavity is to turn  $H_{\rm free}$ into $H_{\rm eff}=H_{\rm ren}(\omega_c)$, and the cavity-induced effect is measured by the difference $H_{\rm eff}-H_{\rm free}$. For example, a small Fabry-P\'{e}rot cavity ($L_z\to0$) formally shifts the modes to higher energy, thus making them less relevant for a modification of the material properties at low energies. In this high frequency limit,  the effect of the Fabry-P\'{e}rot cavity can therefore be considered as an {\em undressing} rather than a {\em dressing} of the electrons with light. 

From a theoretical perspective, both situations pose the same challenge of computing a dressed Hamiltonian $H_{\rm eff}$ for an extended lattice system coupled to a continuum of cavity photon modes, starting from either $H_0=H_{\rm free}$ or from $H_0=H_{\rm bare}$. In this article, we approach this problem in the limit of large cavity frequencies $\omega_c\gtrsim W$, where $W$ is a relevant electronic scale such as the single-particle bandwidth. We develop a high-frequency expansion in $1/\omega_c$ of the effective Hamiltonian, which is nonperturbative in the coupling strength. The formalism can be generically applied to both Situation I and II, as well as different cavity structures. We first study the hybrid light-matter states in Situation I, where the coupling to a few cavity modes is
strongly enhanced \cite{forn-diaz2019,kockum2019}, and demonstrate the formation of a photon-dressed band structure and superconducting pairing due to photon-mediated interactions. In these calculations, a phenomenological coupling parameter $g_{\rm eff}$ is varied to explore different parameter regimes. We then briefly consider Situation II in a conventional Fabry-P\'{e}rot cavity, and estimate the strength of the cavity-induced effects. 

The article is organized as follows. Sec.~\ref{sec:method} discusses the general formalism of lattice electrons coupled to a continuum of electromagnetic modes. The Brillouin-Wigner perturbation theory is introduced to obtain the high-frequency effective model in Sec.~\ref{sec:hfreq}. Sec.~\ref{sec:band} discusses the quantum electrodynamical engineering of electronic bands such as the shape of Fermi surface and the band topology. Sec.~\ref{sec:int} then considers the interaction mediated by photons, which leads to long-range paired states. Sec.~\ref{sec:FP} considers a distinct scenario corresponding to a Fabry-P\'{e}rot cavity. Sec.~\ref{sec:con} provides a conclusion and outlook.

\section{Formulation}
\label{sec:method}
We consider a single-orbital lattice electronic system coupled to a general set of cavity modes through the Peierls-phase coupling, which explicitly preserves gauge invariance \cite{li2020}. The cavity is extended in $x$ and $y$ direction, i.e., $L_x,L_y\to\infty$, so that the modes form a continuum in the momentum space, and can be parametrized by $xy$--momentum $\bm q=(q_x,q_y)$. The Hamiltonian then reads
\begin{align}
\label{ham}
	\hat{H} = \sum_{ ij}\hat{h}_{ij} e^{i\chi_{ij}} + \sum_{\lambda \bm q} \omega_{\bm q} a_{\bm q\lambda}^\dag a_{\bm q\lambda},
\end{align}
where $\hat{h}_{ij}=t_{ij}c^\dag_i c_j $ is the hopping operator from site $i$ to $j$, with $t_{ij}$ the hopping integral and $c_i$ annihilating an electron at site $i$. The photon modes, annihilated by $a_{\bm q}$, are labelled by $\bm q$ and other relevant quantum number $\lambda$. For example, $\lambda$ can include polarization and/or the $z$--momentum $q_z$ of the photon mode. Electrons are coupled to the modes through 
the Peierls phase 
\begin{align}
\chi_{ij}\approx e\bm A\left(\bm R_{ij}\right)\cdot \bm d_{ij},
\end{align}
where $e$ is the electron charge,  $\bm d_{ij}=\bm R_i-\bm R_j$  the bond vector, and $\bm R_{ij}=\frac{\bm R_i+\bm R_j}{2}$. We have assumed the vector potential does not vary strongly over the size of a unit cell, which is generally satisfied by the photon modes in a microwave or optical cavity. We also ignore the possibility of spontaneous symmetry breaking in the photon state (superradiance). Physically, the momentum $\bm q$ of the photon mode should always be restricted in the first Brillouin zone of the electronic lattice model, and should therefore be understood as a \emph{lattice gauge field} in real space. In practice, photon momenta $\bm q$ outside the first Brillouin zone correspond to energies far beyond what is affected by a cavity, so that their consideration is not really needed to understand cavity-induced changes in the materials. Assuming the optical medium is homogeneous inside the material, we can expand the quantized vector potential on the plane-wave basis \cite{glauber1991}, 
 \begin{align}
\bm A(\bm  R)=\sum_{\bm q\lambda} (\mathcal{A}_{\bm  q\lambda}\bm{e}_{\bm q\lambda}a_{\bm  q\lambda} e^{i\bm{ q}\cdot\bm R}+\mathcal{A}_{\bm  q\lambda}^*\bm{e}_{\bm q\lambda}^*a_{\bm  q\lambda}^\dag e^{-i\bm{ q}\cdot\bm R}),
 \end{align} 
 with polarization $\bm e_{\bm q\lambda}$. To simplify the notation, we will not explicitly show the subscript $\lambda$ in the following; it can be readily added back in all equations. The normalization condition for the mode wave functions then requires the coupling to each mode to satisfy $\mathcal{A}_{\bm q}\propto 1/\sqrt{L_xL_y}$. This scaling is crucial for establishing the high-frequency expansion below. 
For later convenience, we recast the 
Peierls phase in the form
 \begin{align}
\chi_{ij}=\sum_{\bm q} g_{ij}({\bm  q})a_{\bm  q} e^{i\bm{ q}\cdot\bm R_{ij}}+H.c.,
 \end{align} 
where the relevant coupling constant is written as  
\begin{align}
\label{gamma}
g_{ij}({\bm  q})=e \mathcal{A}_{\bm  q}\bm{e}_{\bm q} \cdot {\bm d}_{ij} \equiv 2\pi \gamma_{ij}({\bm  q})d/\sqrt{L_xL_y}, 
\end{align} 
where $\gamma_{ij}(\bm q)$ is a dimensionless constant, and $d$ is a length scale of the orders of lattice constant. 

To this point, we have not made any specific assumption on the cavity geometry and the spatial profile of the cavity modes (apart from the validity of the dipolar approximation explained after Eq.~(2)), and the mode structure is fully encoded in the momentum-dependent coupling constant $g_{\bm q}$. This will allow us to develop a general framework treating different kinds of cavities with multiple photon modes. We have assumed the one-band model for simplicity. This would directly apply to materials which can be described by single-orbital lattice models, particularly when the cavity frequencies are off-resonant with the energy gaps. In general, multiple electronic energy bands can be relevant in realistic systems, which can be treated in a similar manner as in this paper (in this case, the light-matter coupling will include also dipolar inter-band matrix elements.) 

\subsection{The quantum Floquet formalism}
As discussed above, it is crucial to precisely treat the Peierls-phase coupling to avoid breaking gauge invariance. Since the Peierls phase factor contains terms of arbitrarily high powers in the vector potential $\mathcal{A}_{\bm q\lambda}$, preservation of gauge invariance therefore requires a nonperturbative treatment of the light-matter coupling. This is particularly important for the strong light-matter coupling regime, where a simple truncation of the Peierls phase can break important physical requirements \cite{li2020,schueler2021}. For this purpose, we will adopt a recently developed quantum Floquet formalism. Specifically, we will expand the Hamiltonian on the basis of photon Fock states \cite{schaefer2018} and subsequently integrate out the photon degrees of freedom using a high-frequency expansion. This procedure generates results that are perturbative in $1/\omega_c$ but \emph{nonperturbative} in the coupling strength, which is suitable for treating strongly coupled light-matter systems.

The formalism has been previously established for lattice systems coupled to a single photon mode \cite{sentef2020,li2020prl}, with emphasis on the analog to the Floquet formalism \cite{mikami2016}. In this article, we generalize the method to a continuum of photon modes. We use the label $\bm n = \begin{pmatrix} n_1, n_2, \ldots, n_M\end{pmatrix}$ to parametrize the Fock state $\ket{{\bm n}}$ of the cavity with $M$ modes, where $n_i$ is the photon number of the $i$th mode. The Hamiltonian is then expanded as 
\begin{align}
\label{mayrex}
\hat{H}=\sum \bra{{\bm n}} \hat{H}\ket{\bm m} (\ket{\bm n}\bra{\bm m}\otimes \mathbb{I}_{\rm el}), 
\end{align}
where $\mathbb{I}_{\rm el}$ is the identity operator for the electron sector. We define the generalized quantum Floquet matrix $\mathcal{H}_{\bm n, \bm m}=\bra{\bm n}\hat{H}\ket{\bm m}- \delta_{\bm n \bm m}\sum_{\bm  q} n_{\bm  q} \omega_{\bm  q} $.
To evaluate $\mathcal{H}_{\bm n, \bm m}$, we note that the matrix element of the Peierls phase operator $e^{i\chi_{ij}}$ factorizes over the different modes $\bm q$, where each factor has the same form as obtained in the single mode case \cite{li2020prl}, so that $\mathcal{H}_{\bm n, \bm m}$ can be written in the form
\begin{align}
\mathcal{H}_{\bm n, \bm m}= \sum_{\langle ij\rangle} \hat{h}_{ij} \prod_{\bm  q} i^{|n_{\bm  q}-m_{\bm  q}|} e^{i(n_{\bm  q}-m_{\bm  q})\phi_{ij}(\bm  q)} j_{n_{\bm  q},m_{\bm  q}}(2|g_{ij}(\bm q)|).
\label{Hmm}
\end{align}
Here $\phi_{ij}(\bm q)=\eta_{ij}(\bm q)+\bm{ q}\cdot\bm{R}_{ij}$, where $\eta_{ij}(\bm q)$ is the phase of the dimensionless coupling $\gamma_{ij}(\bm q)$, $\gamma_{ij}=|\gamma_{ij}|e^{i\eta_{ij}}$. By normal ordering the Peierls phase factor before evaluating the expectation value \cite{li2020prl}, the remaining  $j_{n_{\bm  q},m_{\bm  q}}(2|g_{ij}(\bm q)|)$ is written as the product of an exponential  factor $e^{-|g_{ij} (\bm q)|^2/2}$ and a finite polynomial in $|g_{ij} (\bm q)|$ (see appendix for the precise form).

The matrix representation \eqref{mayrex} has a clear physical interpretation which establishes a connection with the Floquet formalism for laser-dressed states in solids \cite{kitagawa2011}. By expanding the Hamiltonian on the cavity Fock states, we have effectively decomposed the many-body Hilbert space into photon-number sectors labelled by $\bm n$, each of which is described by the quantum Floquet Hamiltonian $\mathcal{H}_{\bm n, \bm n}$. Different sectors are coupled through the off-diagonal matrix elements $\mathcal{H}_{\bm n,\bm m}$. When $\omega_{\bm  q}$ becomes small the electrons can easily excite plenty of infrared photons, and nearby photon sectors strongly mix up. This problem is resolved in a cavity where the frequencies have an infrared cutoff $\omega_c$ determined by the cavity geometry. Emission of real photons with $\omega_{\bm  q}\ge  \omega_c$ then requires a finite energy cost, and the transition between different photon number sectors is suppressed by the cutoff energy $\omega_c$. This suggests a systematic study of the high-frequency limit analogous to the Floquet formalism \cite{mikami2016}, where an effective Hamiltonian is obtained by downfolding the quantum Floquet Hamiltonian \eqref{Hmm} to the zero-photon sector. 
(This high-frequency description is analogous to the stroboscopic motion of periodically driven systems \cite{bukov2015,sentef2020}.) 

In the following we apply the Brillouin-Wigner perturbation theory to integrate out the photon modes, in analogy to the Floquet formalism. More details for the expansion procedure can be found in Ref.~\citenum{mikami2016}. For a given photon-number sector labelled by $\bm n$, the effective Hamiltonian reads
\begin{align}
H_{{\rm eff},\bm n}=\mathcal{H}_{\bm n,\bm n}-\sum_{ \bm m \ne \bm n}\frac{\mathcal{H}_{\bm n, \bm m}\mathcal{H}_{\bm m, \bm n}}{(\bm m - \bm n) \cdot \bm  \omega}+\ldots,
\label{BW}
\end{align}
with the inner product $(\bm m - \bm n) \cdot \bm  \omega=\sum_{\bm  q} (m_{\bm  q}-n_{\bm  q}) \omega_{\bm  q}$. In the following, we will evaluate this series up to the first-order correction, where the zeroth order gives rise to a renormalization of the electron hopping, and the first order leads to induced longer range hoppings and long range interactions.  We note that this high-frequency expansion can also be evaluated for small cavities, where $L_x, L_y$ are of similar orders of magnitudes as $L_z$. More details can be found in Ref.~\citenum{li2020prl}. In the following, we will concentrate on large systems where $L_{x,y}\to\infty$. 

\subsection{Band renormalization}
We first consider the zeroth-order effective Hamiltonian according to Eq.~\eqref{BW},
\begin{align}
H^{(0)}_{{\rm eff}, \bm n}=\mathcal{H}_{\bm n,\bm n}= \sum_{ij} r_{ij} \hat{h}_{ij} ,
\end{align}
in which the electron hopping has been renormalized by the factor $r_{ij}=\prod_{\bm  q} j_{n_{\bm  q},n_{\bm  q}}(2|g_{ij}(\bm q)|)$. To evaluate the renormalization factor, we note that 
\begin{align}
\label{expansionjnn}
j_{n,n}(2|g_{ij}|) = e^{-|g_{ij}|^2/2}(1-n|g_{ij}|^2+\mathcal{O}(|g|^4)), 
\end{align}
and compute the product over $\bm q$ order-by-order in $g$. 
The evaluation can be significantly simplified with a power-counting argument in the limit $L_x,L_y\to\infty$: A term with $|g|^{2k}\sim 1/(L_xL_y)^{k}$ needs $k$ independent momentum summations to yield nonzero results for $L_xL_y\to \infty$. Therefore, terms in the expansion \eqref{expansionjnn} with orders higher than two do not contribute at all in this limit, and the remaining terms can be exactly resummed in the exponent (see appendix for more details)
\begin{align}
\prod_{\bm  q} j_{n_{\bm  q},n_{\bm  q}}(2|g_{ij}(\bm q)|)
=
e^{-\sum_{\bm q} |g_{ij}(\bm q)|^2(n_{\bm q}+\frac{1}{2})}.
\end{align}
Finally we replace $\frac{1}{L_xL_y}\sum_{q_x, q_y}\to \int^\Lambda d^2\bm q$, where the integral $\int^\Lambda d^2\bm q  =\int_{|\bm q |<\Lambda} dq_x dq_y$ is truncated at some ultraviolet cutoff $\Lambda$. Using the parametrization \eqref{gamma} of the couplings, one obtains 
$r_{ij}=\exp[{-d^2  \int^\Lambda d^2\bm q  \left(n_{\bm q }+\frac{1}{2}\right)|\gamma_{ij}(\bm  q)|^2}]$. The sum over the different photon branches $\lambda$, labelling polarization and $z$-quantization, could be easily reinstated  in the exponent. The low energy sector is then simply obtained by setting $n_{\bm  q}=0$ in this expression. 
For the specific settings  studied below, we will parametrize the renormalization factor as 
\begin{align}
r_{ij}=e^{-d^2\int^\Lambda d^2\bm q  \frac{1}{2}|\gamma_{ij}(\bm  q )|^2} \equiv e^{-\pi z_{ij} g_{\rm eff}^2},
\label{rij}
\end{align} 
where $g_{\rm eff}$ quantifies the overall coupling strength, and the bond factor $z_{ij}$ is a geometry dependent factor of order one. To be concrete, we can always decompose $\gamma_{ij}=\gamma \theta_{ij}$ where $\theta_{ij}$ captures all bond-dependence, so that $z_{ij}=\int^\Lambda d^2\bm q   \frac{1}{2}|\gamma_{ij}(\bm  q )|^2/\int^\Lambda d^2\bm q \frac{1}{2}|\gamma(\bm  q )|^2$. The functional form of $\theta_{ij}$ depends on the geometry of the cavity and the lattice system, which will be explicitly given in specific situations. In the following practical calculation, we will always assume that $\theta_{ij}$ only depends on the direction of $\bm q$, and the bond factor can be recast into the simple form 
\begin{align}
z_{ij}=\int^{2\pi}_0\frac{d\varphi}{2\pi}|\theta_{ij}(\varphi)|^2,
\end{align}
where $\varphi$ is the polar coordinate of $\bm q$.

Eq.~\eqref{rij} shows the advantage of the high-frequency expansion, where the coupling can be treated nonperturbatively: The light-induced bandwidth suppression $r_{ij}$ emerges naturally from the nonlinear gauge-invariant coupling and includes terms of all orders in $g_{\rm eff}$.  

To provide a concrete example, we estimate the renormalization due to the coupling to a given photon branch in a Fabry-P\'{e}rot cavity consisting of two metallic mirrors separated by $L_z$. The relevant modes with  electric field in the direction of the material are the transverse magnetic (TM) modes of the Fabry-P\'{e}rot cavity. We take the lowest mode with dispersion $\omega_{\bm q }=v\sqrt{q_c^2+\bm q^2}$, where $q_c=\omega_c/v=\pi/L_z$ is the cavity momentum. The dimensionless coupling has the form \cite{schlawin2019}
\begin{align}
\label{eq:gamma}
\gamma_{ij}(\bm  q )=\theta_{ij}({\bm q }) \sqrt{\frac{\alpha v}{2\pi \sqrt{\epsilon_r} L_z \omega_{\bm  q}}},
\end{align}
where $\alpha=e^2/4\pi\epsilon_0\hbar c$ is the fine structure constant and $\epsilon_r$ is the relative permittivity. The speed of light in medium $v=c/\sqrt{\epsilon_r}$. It is worth noting that the Fabry-P\'{e}rot cavity generally has a $\theta_{ij}({\bm q})$ depending on the magnitude $|\bm q|$, which will be ignored in calculations to estimate the order of magnitude of the effective coupling $g_{\rm eff}$. With this, we reach the following expression,
\begin{align}
\label{eq:geff}
g_{\rm eff}&= \frac{d}{\sqrt{2\pi}}\left(\int^\Lambda d^2\bm q  |\gamma(\bm q )|^2\right)^{\frac{1}{2}}\nonumber\\
&=\frac{d}{\lambda_c}\sqrt{\frac{2\alpha }{\sqrt{\epsilon_r}}}  \left(\sqrt{1+\frac{\Lambda^2}{q_c^2}}-1\right)^{\frac{1}{2}},
\end{align}
which depends on the geometrical parameters of the cavity and the number of relevant photon modes quantified by the cutoff $\Lambda$. For example, one can consider a simple square lattice with lattice constant $d=0.5$nm and assume $\epsilon_r=1$, $\lambda_c=0.5\mu$m, and $\Lambda=10q_c$, and the coupling strength can be evaluated to be $g_{\rm eff}\approx 4\times 10^{-4}$, similar to the analogous results in atomic systems \cite{devoret2007}. The proper value of the cutoff $\Lambda$, characterizing the ability of confining high-frequency photon modes, depends on the actual structure and electromagnetic properties of the cavity and should be determined using first-principle simulations in the most general settings. It is worth noting that, in free-space solids, the renormalization $r_{ij}$ has already been summed over all modes and absorbed into the hopping parameter $t_{ij}$, while the cavity-induced renormalization corresponds to the change in $r_{ij}$ when the cavity is present. This analysis will be performed in Sec.~\ref{sec:FP}. 

In subwavelength cavities, a further enhancement 
can be gained by 
compression of the effective mode volume $V_{\rm mode}$ 
below 
the volume $V=L_xL_yL_z$ of the whole experimental setup. To strengthen $g_{\rm eff}\sim 10^{-4}$ to the ultrastrong coupling regime $g_{\rm eff}\gtrsim 0.1$, an enhancement factor of $A\sim 10^5$ is necessary \cite{schlawin2019,wang2019}. This setting could then correspond to Situation I discussed in the introduction, where the free-space matter Hamiltonian includes the dressing by vacuum photons, while the renormalization factor related to the compressed modes is treated separately and accounts for the cavity effect on the band structure.

\subsection{The induced hopping and interaction}
\label{sec:hfreq}
One can derive the light-induced Hamiltonian order by order using Eq.~\eqref{BW}. The expansion is controlled by the frequency 
in the denominator. In the following, we will concentrate on the first-order correction for the low-energy sector ($\bm n=\bm0$). The induced Hamiltonian can then be recast into the 
form
\begin{align}
H^{(1)}_{\rm eff}=-\sum_{ \bm l\ne \bm 0}\frac{\mathcal{H}_{\bm 0, \bm l}\mathcal{H}_{\bm l,\bm 0}}{\bm l \cdot \bm  \omega},
\end{align}
where we have omitted the subscript $\bm n=\bm 0$. The factor $\mathcal{H}_{\bm 0, \bm l}\mathcal{H}_{\bm l,\bm 0}$ involves intermediate virtual states labelled by the photon-number vector $\bm l$ and is evaluated as an infinite product of the $j_{0l}$ functions. The calculation can be simplified again using the power counting argument: Intermediate states with multiple photons from the same mode ($ l_i\ge2$) constitute vanishingly small phase space and can be neglected in the continuum-mode limit. It is only possible to absorb/emit many virtual photons from \emph{different} modes. 
Details of the derivation are given in the appendix. 
The result for a single photon branch is summarized as follows,
\begin{align}
H^{(1)}_{\rm eff}&=-\sum_{ijk,\sigma} t_{ij}t_{jk}V^{ij}_{jk}c^\dag_{i\sigma}c_{k\sigma}
-\sum_{ij, i'j'}V^{ij}_{i'j'}\normord{\hat{h}_{ij}\hat{h}_{i'j'}},
\label{hopnint}
\end{align}
where the normal-ordering $\normord{\mathcal{O}}$ with respect to the empty-lattice state moves all annihilation operators in $\mathcal{O}$ to the right. The quadratic induced-hopping term arises due to normal ordering the product $h_{ij}h_{i'j'}$.
If the renormalization due to a single photon branch is considered,
the effective interaction vertex $V^{ij}_{i'j'}$ is explicitly given by
\begin{align}
&V^{ij}_{i'j'}=r_{ij}r_{i'j'}\sum_{l=1}^{\infty}\frac{(-)^{l}d^{2l}}{l!}
\int^\Lambda \Big(\prod_{s=1}^l d^2\bm q_{s}\Big)
\nonumber\\
&
 \times\,\,\,\,e^{-i \sum_{s=1}^{l} [\bm  q_{s}\cdot  (\bm R_{ij}-\bm R_{i'j'})]} \frac{\prod_{s=1}^{l} \gamma^*_{ij}(\bm q_{s})\gamma_{i'j'}(\bm q_{s})}{\sum_{s=1}^{l} \omega_{\bm q_{s}}}.
\label{induced_int}
\end{align}
This is the central result of this article. 

For the coupling to multiple photon branches, one should reinstate the quantum number $\lambda$ for $\gamma_{ij}( \bm q,\lambda)$ and $\omega_{\bm q, \lambda}$ and replace the integration $\int ^\Lambda d^2\bm q\to \sum_{\lambda}\int ^\Lambda d^2\bm q$ (see appendix). The induced Hamiltonian \eqref{hopnint} contains two terms: (i) the induced hopping which consists of two bare hopping processes from $i\to j$ and $j\to k$ accompanied by photon absorption and emission, and (ii), the induced interaction which couples the hopping processes $i\to j$ and $i'\to j'$ at different bonds. The induced hopping clearly modifies the electronic energy band, which will be demonstrated below. 

\section{Cavity control of the band structure}
\label{sec:band}
The formulation carried out in the previous sections is applicable to a general lattice system minimally coupled to a continuum of photon modes with arbitrary dispersion. In this section, we consider the cavity geometry depicted in Fig.~\ref{fig:cavity}, which can be, e.g., a scalable nanoplasmonic cavity with macroscopic sizes in $xy$--directions. 
The couplings to modes
near the lowest frequency $\omega_c$ are 
assumed to be strongly enhanced, as 
described for Situation I in the introduction.
 For the matter part, we study two-dimensional lattices with nearest-neighbor hopping $t_0$. The photon polarization is approximately aligned in the $xy$-plane inside the matter. We will concentrate the cavity-induced effects and ignore the intrinsic Coulomb interactions which are already present in the material even without the cavity. It is important to note that Coulomb repulsion can be incorporated in the high-frequency framework by adding interaction terms in the effective Hamiltonian and can be treated with conventional perturbation theories in the weak-(Coulomb-)coupling limit and with numerical methods, such as dynamical mean-field theory, in the strong-coupling limit. 
 
To proceed with analytical calculations, we assume the form of $\gamma_{ij}$ as given by Eq.~\eqref{eq:gamma} where the factor $\theta_{ij}$ only depends on the direction of $\bm q$. It is natural to assume the UV cutoff $\Lambda$ is much smaller than the crystal momentum $\sim 1/ |\bm r_{ij}|$, since the latter generally corresponds to X-ray photon energy well beyond the electronic band around Fermi level. To be general, we further include the different polarization modes labelled by $\lambda=1,2$, which have degenerate frequencies for given momentum, $\omega_{\bm q\lambda}=\omega_{\bm q}$. In this case, the induced hopping term (the first term in $ H_{\rm eff,1}$) can be written as $-\sum_{ij,\sigma}t_{ij}^{\rm ind}c^\dag_{i\sigma} c_{j\sigma}$ with hopping amplitudes
\begin{align}
t_{ij}^{\rm ind}=\sum_{k}\frac{1}{\omega_c}t_{ik}t_{kj}r_{ik}r_{kj}f\left(-\sum_\lambda z_{ik,kj}(\lambda)g^2_{\rm eff, \lambda}\right),  
\label{eq:hopping}
\end{align}
where we have defined the following function $f(x)=2\pi x\log(\sqrt{1+\Lambda^2/q_c^2})/(\sqrt{1+\Lambda^2/q_c^2}-1)+\mathcal{O}(x^2)$, which is obtained by evaluation of the integral \eqref{induced_int}, and the coupling $g_{\rm eff}$ is given by Eq.~\eqref{eq:geff} (see appendix \ref{appB} for details). The summation over polarization $\lambda$ has been conducted inside the $f(x)$ function, and the two-bond factor $z_{ik,kj}$ comes from the angular integration including $\theta_{ik}\theta_{kj}$. Specifically, since we assumed $\theta_{ik}=\theta_{ik}(\varphi)$ where $\varphi$ is the angular coordinate of $\bm q$, the two-bond factors read 
\begin{align}
z_{ik,kj}(\lambda)=\int \frac{d\varphi}{2\pi} \theta^*_{ik}(\lambda,\varphi)\theta_{kj}(\lambda,\varphi).
\end{align}
In the following,  $g_{\rm eff}$ will be treated as a phenomenological parameter and will be varied to explore different parameter regimes.

\subsection{Band renormalization}
\label{2dcavity}
\begin{figure}
\includegraphics[scale=0.35]{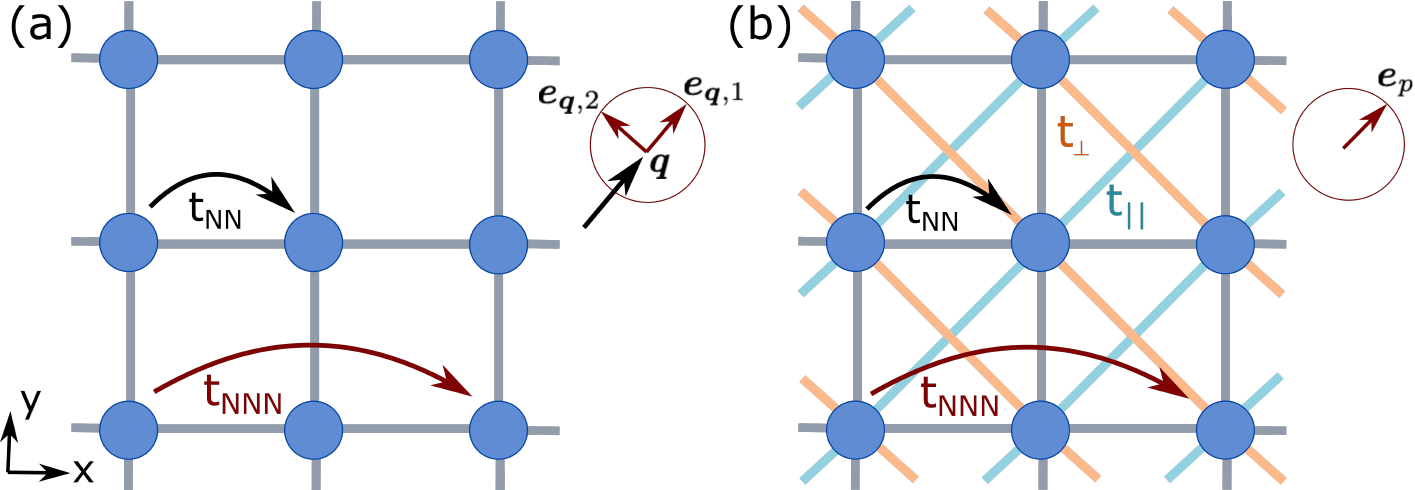}
\caption{The square-lattice system with cavity-induced electronic hopping. The original system features a nearest-neighbor hopping $t_{\rm NN}=t_0$. (a) System placed in an isotropic cavity with two polarization modes per each $\bm q$. A nearest-neighbor hopping $t_{\rm NNN}$ is induced while the four-fold rotational symmetry is preserved. (b) System placed in a fixed-polarization cavity with the mode polarization always along the diagonal direction. The setup gives rise to induced hopping along diagonal (indigo, $t_{\parallel}$) and anti-diagonal (orange, $t_{\perp}$) directions.}
\label{fig:lat}
\end{figure}

We first consider a square-lattice system placed in an isotropic 2D cavity (see Fig.~\ref{fig:lat}). Two modes are included for each momentum $\bm q=(q_x,q_y)$, whose polarization direction is inspired by the modes in the Fabry-P\'{e}rot cavity: one with polarization $\bm e_{1,\bm q}= \bm q/|\bm q|$ and the other with $\bm e_{2,\bm q}=\bm e_{1,\bm q}\times \hat{\bm z}$ \cite{kakazu1994}. A schematic depiction can be found in Fig.~\ref{fig:lat}(a). We consider the following simplest form of the bond dependence factors.
\begin{align}
\label{eq:pseudofp}
\theta_{1,x}(\bm q)=\theta_{2,y}(\bm q)=\cos\varphi,\nonumber\\ \theta_{1,y}(\bm q)=-\theta_{2,x}(\bm q)=\sin\varphi,
\end{align}
where $\varphi$ is the angular coordinate of $\bm q$. 
\begin{figure}
\includegraphics[scale=0.7]{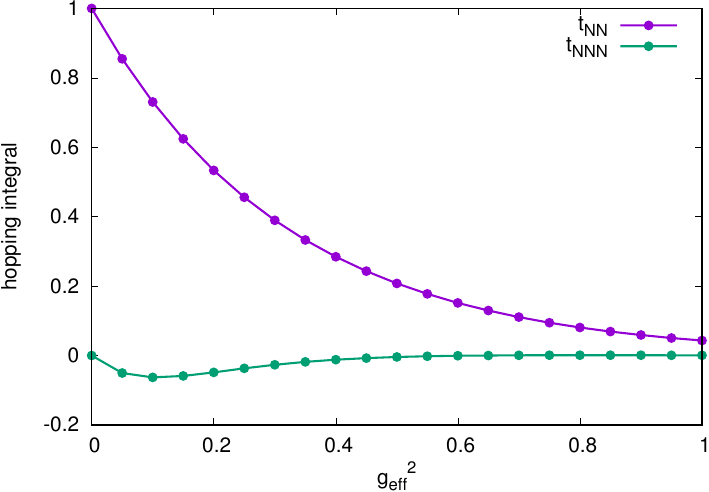}
\caption{The hopping integral for the effective Hamiltonian in the isotropic 2D cavity, with the theta factors as given in \eqref{eq:pseudofp}. For simplicity, the overall factor $f_{\rm eff}$ is set to $1$. $t_{\rm NN}$ and $t_{\rm NNN}$ represent nearest-neighbor and next-nearest-neighbor hopping amplitude along $x$-- or $y$--directions, respectively.}
\label{fig:hop_FP}
\end{figure}
\begin{figure}
\includegraphics[scale=1.8]{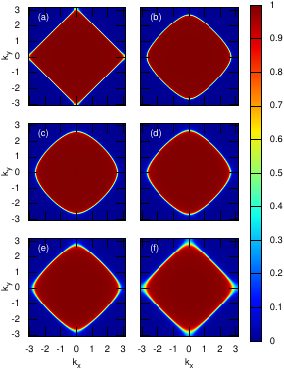}
\caption{The non-interacting band occupation of the square-lattice solid inside a Fabry-P\'{e}rot cavity. The system is half-filled ($\mu=0$). The panels (a)-(f) correspond to $g^2_{\rm eff}=0.0,0.05,0.10,\ldots,0.50$.}
\label{fig:nk_FP}
\end{figure}
For the hopping modification, the first effect is the renormalization $t_0\to  t_0e^{-\pi g^2_{{\rm eff}}}$. The bandwidth shrinks as $g^2_{\rm eff}$ increases. A quasi-flat energy band emerges for very large couplings. Another effect is the induced hopping, which is characterized by the interaction vertex $V^{ij}_{jk}$ for neighboring bonds sharing a common lattice site $j$. The main feature of this setup is its isotropy in the $xy$-plane, so that the 4-fold rotational symmetry of the square lattice is preserved. In this case, the bond factors of both polarizations are identically $z_{xy}=z_{yx}=0$ and $z_{xx}=z_{yy}=1/2$ for the nearest-neighbor bonds along $x,y$-directions. Using the above $z$-factors (see appendix), 
we have
\begin{align}
t_{\rm NNN}= \frac{t_0^2}{\omega_c}e^{-2\pi g^2_{{\rm eff}}}f(-g^2_{\rm eff}),
\end{align}
for next-nearest-neighbor hopping along $x$-- and $y$--directions. In the following, we formally set $t_0^2/\omega_c=1$ and assume $\Lambda=10q_c$ unless otherwise stated. 

We show the cavity-modified hopping amplitudes in Fig.~\ref{fig:hop_FP}. The original hopping $t_0$ is renormalized to $t_{\rm NN}$ as the light-matter coupling is turned on, and a next-nearest-neighbor hopping $t_{\rm NNN}$ emerges. Both hopping parameters decay in the strong coupling regime, which is attributed to the renormalization factor $e^{-\pi g^2_{{\rm eff}}}$. Fig.~\ref{fig:nk_FP} shows the electron occupation in the momentum space for different coupling strengths (ignoring the cavity-induced interactions). The modified hopping leads to deformation of the Fermi sea as the light-matter coupling $g_{\rm eff}^2$ increases. 

\subsection{Cavity-induced dimensional reduction}
\begin{figure}
\includegraphics[scale=0.7]{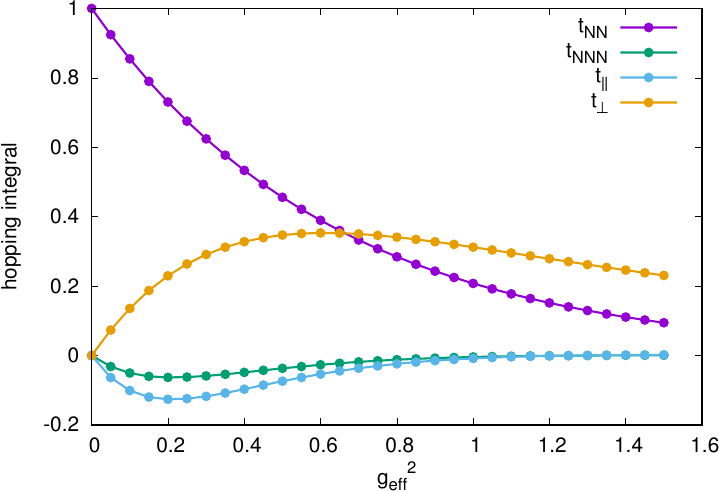}
\caption{The hopping integral for the effective Hamiltonian in the fixed-polarization cavity. The overall scale is again set to one. $t_{\rm NN}$ and $t_{\rm NNN}$ again indicate nearest-neighbor and next-nearest-neighbor hopping amplitude along $x$-- or $y$--directions. $t_{\parallel}$ represents hopping along $(1,1)$ direction, which is parallel to the cavity polarization; $t_{\perp}$ represents hopping along $1,-1$ direction, which perpendicular to the polarization.}
\label{fig:hop_line}
\end{figure}
\begin{figure}
\includegraphics[scale=1.8]{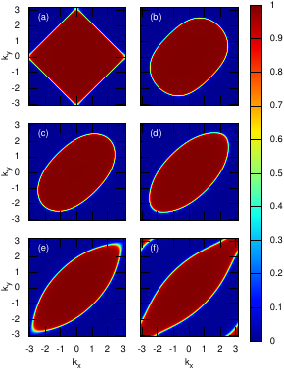}
\caption{The non-interacting band occupation of the square-lattice solid inside the fixed-polarization cavity. The system is half-filled ($\mu=0$). The panels (a)-(f) correspond to $g^2_{\rm eff}=0.0,0.30,0.60,\ldots,1.50$.}
\label{fig:nk_line}
\end{figure}

We next consider a different scenario by assuming the polarization is fixed for all photon modes in the cavity, which will be referred to as a fixed-polarization cavity in the following discussion. This is a representative case for cavities breaking the crystal symmetry of the lattice. The specific type of cavity can be realized by, e.g., making $L_y$ as small as the cavity wavelength $\lambda_c=2\pi v/\omega_c$ while keeping $L_x$ macroscopic. In this way, the photon momentum in $y$--direction becomes discrete and one can further assume only one $q_y=2\pi/L_y$ is dominant. The material can still be treated as a two-dimensional lattice if $L_y\sim \lambda_c$ is much larger than the lattice constant. This setup is related to a tranmission-line cavity \cite{hagenmueller2010}.  

For simplicity of calculations, we assume the momentum $\bm q$ still takes values from the two-dimensional phase space but the polarization $\bm e_{\bm q}$ is fixed to the diagonal direction of the square lattice, see Fig.~\ref{fig:lat}(b). This cavity setup leads to fixed $\theta_x=\theta_y=1/\sqrt{2}$ for all momenta $\bm q$ and bond factors $z_{xx}=z_{yy}=z_{xy}=z_{yx}=-z_{x,-y}=-z_{-x,y}=1/2$. The original nearest-neighbor hopping $t_0$ again gets renormalized $t_0\to t_0 e^{-\pi g_{\rm eff}^2/2}$ and a next-nearest-neighbor $t_{\rm NNN}=\frac{t_0^2}{\omega_c}e^{-\pi g^2_{{\rm eff}}}f(-g^2_{\rm eff}/2)$ emerges. In addition, the cavity induces hopping processes along the diagonal and anti-diagonal directions. These induced hoppings explicitly break the 4-fold rotational symmetry of the free-space matter Hamiltonian. Using the above $z$-factors, we obtain
\begin{align}
t_{\parallel}&=\frac{t_0^2}{\omega_c} e^{-\pi g_{\rm eff}^2}f(-g_{\rm eff}^2/2)\nonumber\\
t_{\perp}&=\frac{t_0^2}{\omega_c} e^{-\pi g_{\rm eff}^2}f(g_{\rm eff}^2/2).
\end{align} 
The resulting hopping amplitude is shown in Fig.~\ref{fig:hop_line}. For smaller coupling strength, the hoppings $t_{\perp}$ and $t_{\parallel}$ are opposite in sign but approximately the same in magnitude. In the strong coupling regime, however, $t_{\parallel}$ along the diagonal direction becomes significantly suppressed due to the renormalization factor, while $t_{\perp}$ along the anti-diagonal direction becomes dominant. Physically speaking, we find the hopping process \emph{perpendicular} to the fixed polarization does not get strongly suppressed. This seemingly surprising observation can be explained by the fact that the enhancement of $f(g_{\rm eff}^2/2)$ in $t_{\perp}$ (relative to $f(-g_{\rm eff}^2/2)$ in $t_{\parallel}$) plays against the suppression of the renormalization factor $e^{-2\pi g_{\rm eff}^2}$ as $g_{\rm eff}$ increases. It is worth noting that this anti-diagonal hopping is indeed induced by two photon-dressed hoppings along two original bonds, which are neither perpendicular to the cavity polarization. The dimensional crossover is a direct result of the renormalization factor $\exp(-2\pi g_{\rm eff}^2)$ which rises due to the full gauge-invariant Peierls-phase coupling. This is another example where a full treatment of the nonlinear Peierls phase is crucial.
 
An intriguing consequence of this observation is that the lattice system becomes quasi-one-dimensional in the strong coupling regime. The emergence of a one-dimensional system is demonstrated by the change of the Fermi surface, as shown in Fig.~\ref{fig:nk_line}. As $g_{\rm eff}^2$ increases, the Fermi surface deforms from the diamond shape to a stripe-like distribution in $(1,1)$ direction of the reciprocal space. The 
process leads to a Lifshitz phase transition in the sense that the two ends of the Fermi surface touch at the momentum $(\pi,\pi)$  and winds around the torus-shaped first Brillouin zone (FBZ). We finally note that, although the anisotropic band renormalization can be a general phenomenon when the cavity breaks crystal symmetries, such a large effective coupling considered in our calculation is generally unrealistic for an optical cavity (such as Fabry-P\'{e}rot cavity). 

It is interesting to compare the dimensional reduction observed above to the conceptually related phenomenon induced by a DC-electric field \cite{aron2012}. We note that the cavity-induced dimensional crossover does not rely on the presence of a strong Coulomb electric field, but is a result of dressing by strongly coupled transverse photons. 

\subsection{Manipulating band topology}
In this section, we concentrate on the topological properties of the Bloch wave functions, which can naturally be modified by the induced hopping processes. We consider a paradigmatic example, the anomalous quantum Hall effect on a honeycomb lattice, which has been experimentally implemented in cold-atom system \cite{jotzu2014} and laser-driven graphene \cite{mciver2020}. Using the quantum light coupling, we explore an equilibrium implementation of the model. To be concrete, we suppose the cavity is designed so that a single circularly polarized mode with fixed handedness is retained for each momentum $\bm q$, termed a chiral cavity. A chiral cavity can be constructed, e.g., using two parallel Faraday mirrors which consist of a normal mirror and a Faraday rotator \cite{huebener2020}. The key idea is to break the time-reversal symmetry to alter the topological class of the lattice system \cite{kitaev2009,claassen2017}. Similar scenarios have been discussed for smaller systems coupled to one or two photon modes using perturbative methods \cite{wang2019,tokatly2021}. Here we consider the coupling to a continuum of modes in the thermodynamic limit and do not assume a weak effective coupling. 

\begin{figure}
\includegraphics[scale=1.5]{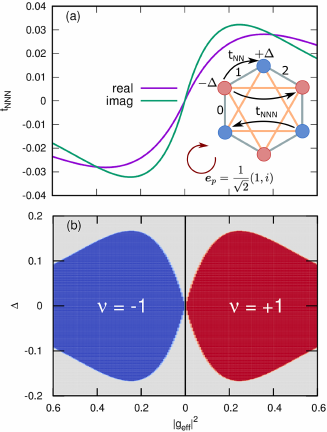}
\caption{Chern insulating phase of a honeycomb lattice system placed in a circularly polarized cavity. (a) The next-to-nearest-neighbor hopping induced by strong light-matter coupling. The inset shows a schematic diagram of a hexagon in the lattice with $t_{NN}$ and $t_{NNN}$ labelled by bonds of different colors. The three distinct nearest-neighbor bonds are labelled by $0,1,2$ as discussed in the main text. Atoms in the two sublattices have different onsite energies $\pm\Delta$ and are shown with blue and red colors. (b) Chern number as a function of effective coupling $\Delta$. The left and right sub-panels correspond to opposite chiralities of the cavity modes. In the grey area Chern number is zero. }
\label{fig:haldane}
\end{figure}

Specifically, we consider the honeycomb lattice with nearest-neighbor hopping $t_0$. For the two sublattices, we suppose an onsite energy $\pm \Delta$, respectively. In this case, the electronic Hamiltonian originally contains hopping of amplitude $t_0$ along three bonds: $(0,1)$, $(\frac{\sqrt{3}}{2},\frac{1}{2})$, $(\frac{\sqrt{3}}{2},-\frac{1}{2})$, labelled by $\alpha=0,1,2$ in Fig.~\ref{fig:haldane}, respectively. We assume each photon momentum has a single mode with fixed circular polarization $\bm e_p=(1,i)/\sqrt{2}$, leading to $\bm q$-independent $\theta_\alpha=\exp(-i\alpha\pi/3)/\sqrt{2}$. The bond dependence factors are evaluated to be $z_{\alpha\alpha'}=\exp(i(\alpha-\alpha') \pi / 3)/2$. Using the same effective-theory method, we obtain the induced next-nearest-neighbor hopping
\begin{align}
t_{\rm NNN}\approx \frac{t_0^2}{\omega_c}e^{-\pi g^2_{{\rm eff}}}f(-e^{i\pi/3}g^2_{\rm eff}/2).
\label{hopchiral}
\end{align}
The direction of hopping is labelled in the inset of Fig.~\ref{fig:haldane}(a). The induced $t_{\rm NNN}$ gives rise to a Haldane model on the honeycomb lattice \cite{haldane1988}. The hopping amplitude is shown in Fig.~\ref{fig:haldane}(a). The induced complex hopping amplitude rises and then gets suppressed as the effective coupling increases. 

In this effective model, an electron captures a finite Berry phase determined by $t_{\rm NNN}$ when it hops around a full loop of an orange triangle in Fig.~\ref{fig:haldane}(a). The two orange triangles in the diagram are related by a lattice reflection.  As a result, this leads to mass gaps $\Delta'_{\tau}=-\tau 3\sqrt{3}\operatorname{Im}t_{\rm NNN}$ of opposite signs for the two distinct Dirac cones with $\tau=\pm1$. When the magnitude of $\Delta'_\tau$ exceeds the trivial gap $ \Delta$, the system transits to a topological phase characterized by the Chern number $\nu=-[\operatorname{sgn}(\Delta+\Delta'_\tau)-\operatorname{sgn}(\Delta-\Delta'_\tau)]/2$, which is plotted in Fig.~\ref{fig:haldane}(b). The left subpanel shows results of the opposite cavity polarization ($e^{i\pi/3}\to e^{-i\pi/3}$ in \eqref{hopchiral}). We see a broad range where the Chern number becomes nonzero ($\pm1$), indicating the emergence of the anomalous quantum Hall effect. As expected, the Chern number changes its sign when the polarization is reversed. The results are consistent with previous works treating the lattice coupled to a single photon mode \cite{wang2019}. 

\section{Photon-mediated interaction}
\label{sec:int}
So far we have concentrated on the induced hopping and ignored other interaction effects. However, the photon-mediated interaction 
is of similar magnitude as the induced hopping. 
Indeed, the interaction is closely related to the possible superradiant phase transition in strongly coupled light-matter system, in analogy to the all-to-all interaction in Dicke model. Moreover, for 
the free electron gas coupled to cavity photon modes, it has been argued that this interaction couples electric currents at different positions, inducing a pair-density-wave (PDW) phase \cite{schlawin2019}. In the following, we systematically discuss different types of emerging interactions due to the gauge-invariant light-matter coupling. In particular, a different type of interaction emerges when higher order terms in the Peierls coupling are considered, which is analogous to the $A^2$ term in the gauge-invariant version of Dicke model \cite{rzazewski1975,knight1978, nataf2010, viehmann2011, todorov2014, todorov2015, bernardis2018, schuler2020}. This interaction, later termed kinetic-type interaction, favors a nonlocal but uniformly paired superconducting state which is distinct from the PDW phase and introduces further modification to the Fermi surface. 

In the following we will discuss the form of the interaction and analyze its physical effects using mean-field theory. We nevertheless stress that the interaction strength is in principle as strong as the induced hopping, and fluctuation around the mean-field solution is not necessarily weak.

\subsection{The kinetic-type and current-type interactions}
\label{sec:int}
First of all, we decompose the interaction vertex in \eqref{induced_int} by $V^{ij}_{i'j'}=(V^K)^{ij}_{i'j'}+(V^J)^{ij}_{i'j'}$, where $V^K$ and $V^J$ only contain the even and odd order terms in the $l$ summation, respectively. In other words, $(V^K)^{ij}_{i'j'}$ and $(V^J)^{ij}_{i'j'}$ are the symmetric and antisymmetric parts under the exchange $i\leftrightarrow j$ or $i'\leftrightarrow j'$. The full interaction term now reads 
\begin{align}
&\frac{1}{4}\sum_{ij, i'j'}[-(V^K)^{ij}_{i'j'}\normord{\hat{K}_{ij}\hat{K}_{i'j'}}+(V^J)^{ij}_{i'j'}\normord{\hat{J}_{ij}\hat{J}_{i'j'}}],
\label{ck_form}
\end{align}
where we have defined the kinetic-energy operator $\hat{K}_{ij}=\hat{h}_{ij}+\hat{h}_{ji}$ and the current operator $\hat{J}_{ij}=i(\hat{h}_{ij}-\hat{h}_{ji})$. Normal-ordering has been individually applied to each term like $\hat{h}_{ij}\hat{h}_{i'j'}$. This result is significantly different from the linear-coupling approximation (truncation of the Peierls phase to the first-order term) which breaks the gauge-invariant coupling form \cite{schlawin2019}.  In that case, the light-matter interaction is approximated by $\sim J_{ij} \chi_{ij}$, where $\chi_{ij}$ is the Peierls phase along bond $ij$, and only the lowest-order $V^J$ term ($\propto \gamma(\bm q)^2$) survives. Since the kinetic-type interaction is of order $\gamma^4$, it is completely missing under this approximation. In this sense, the kinetic-type interaction arises due to gauge-invariance in analogy to the diamagnetic term in Dicke model.

\begin{figure}
\includegraphics[scale=0.48]{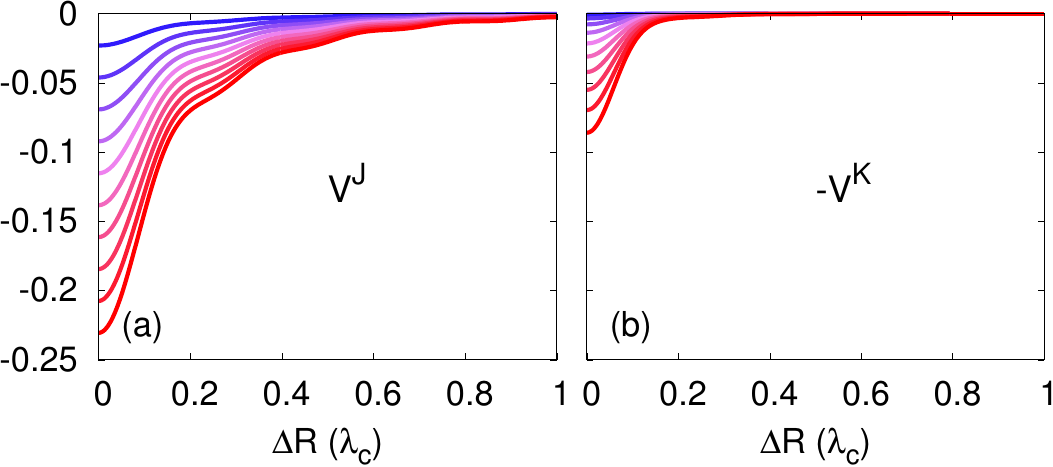}
\caption{The current-type and kinetic-type interactions calculated at the lowest order. The distance is measured in the unit of the cavity wavelength. The curves from blue to red indicate $g^2_{\rm eff}=0.01, 0.02,\ldots, 0.10$. Both curves are rescaled by the same arbitrary factor. Parameters such as $\Lambda$ are the same as before.}
\label{fig:lr}
\end{figure}

To gain some insights on the magnitudes of the two types of interactions, we calculate the lowest-order approximation to $V^K$ and $V^J$ and, for simplicity, ignore the bond dependence. The results are plotted in Fig.~\ref{fig:lr}. It is expected that the interaction range has the orders of magnitude of $\lambda_c$ from the analytic formula \eqref{induced_int}. Both interactions are attractive in the shown range. The current-type interaction is generally stronger than the kinetic-type interaction, since $V^K$ is of $g_{\rm eff}^4$ while $V^J$ is of $g_{\rm eff}^2$. The strengths of the two interactions can nevertheless become comparable in the strong-coupling regime. 

\subsection{Photon-mediated pairing and gapped Fermi surface}
The kinetic-type and current-type interactions have distinct physical properties. In general, the current-type interaction couples the current operators at different bonds and can lead to a novel PDW state with Amperean pairing \cite{lee2014, schlawin2019}. This is because the current-type interaction is attractive for hoppings along the same direction, or electrons of near-parallel momenta in the reciprocal space, in analogy to the Amperean force between parallel currents. In the following, we will show that the kinetic-type interaction can lead to a more conventional BCS pairing state, where electrons with opposite momenta are coupled to form Cooper pairs. The pairing is neverthelss highly nonlocal and is distinct from the conventional $s$-wave pairing. To be concrete, we transform the interaction \eqref{induced_int} to momentum space and consider the mean-field decoupling (Although exponentially decaying, for typical parameters the light-induced interactions are strongly peaked in momentum space, as discussed in the next paragraph. In real space, the interaction therefore  extends over many unit cells, so that the mean-field decoupling is at least a good starting point). Within mean-field decoupling, the Hamiltonian reads
\begin{align}
&\sum_{\bm q\sigma} V_{\bm k,- \bm k, \bm q}c^\dag_{-\bm k-\bm q,\bar{\sigma}} c^\dag_{\bm k +\bm q,\sigma}c_{\bm k\sigma}c_{-\bm k\bar{\sigma}}\nonumber\\
&\quad\to [\sum_{\bm q\sigma}\bar{\Delta}_{\bm k+\bm q} c_{\bm k\sigma}c_{-\bm k\bar{\sigma}}+ \text{H.c.}],
\end{align}
with $\sigma=\uparrow,\downarrow$ and the self-consistent condition $\Delta_{\bm k}=\sum_{\bm q}\bar{V}_{\bm k,- \bm k, \bm q}\langle c_{\bm k+\bm q\uparrow} c_{-\bm k-\bm q \downarrow}\rangle$. The momentum-space interaction vertex $V_{\bm k,\bm k', \bm q}$ is the Fourier transform of $V^{ij}_{i'j'}$ multiplied by a delta function imposing momentum conservation. Note that here $\bm q$ denotes the change in the electronic momentum due to scattering.

\begin{figure}
\includegraphics[scale=0.75]{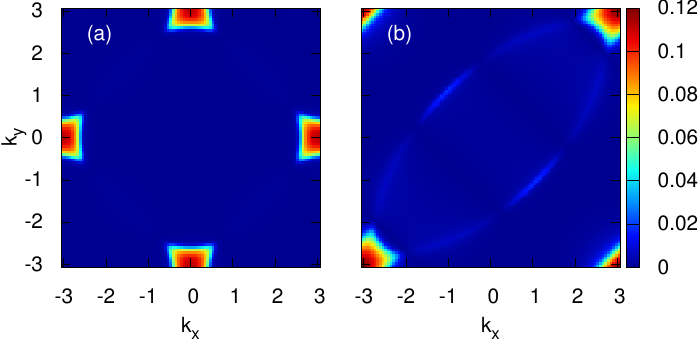}
\caption{The BCS gap parameter $\Delta_{\bm k}$ for the same states plotted in Fig.~\ref{fig:fs}. (a) $g^2_{\rm eff}=0.5$ for a Fabry-P\'{e}rot cavity with 2D dispersion. The superconducting gaps open near the edges of the FBZ around $(\pm \pi,0)$ and $(0,\pm \pi)$. (b) $g^2_{\rm eff}=1.2$ for a fixed-polarization cavity. The gap opens at the corners of the FBZ, i.e., near $(\pi,\pi)$. The inverse temperature $\beta=100$.}
\label{fig:pair}
\end{figure}

\begin{figure}
\includegraphics[scale=0.75]{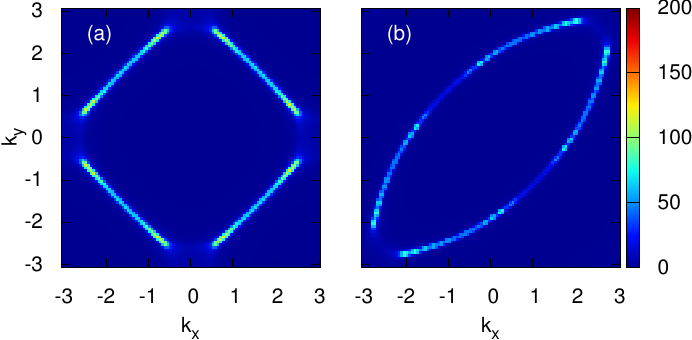}
\caption{The spectral weight at zero energy for the square-lattice solid under strong light-matter coupling. (a) $g^2_{\rm eff}=0.5$ for a Fabry-P\'{e}rot cavity with 2D dispersion.  (b) $g^2_{\rm eff}=1.2$ for a fixed-polarization cavity. The delta function is regularized by a Lorentzian function $ \eta/\pi(\omega^2+\eta^2)$ with $\eta=0.01$. $\beta=100$ as in Fig.~\ref{fig:pair}. }
\label{fig:fs}
\end{figure}

The interaction vertex $V_{\bm k\bm k', \bm q}$ is sharply centered around $\bm q =0$ due to the large speed of light. Indeed, only modes with $|\bm q_{ s}|\ll 1/d$ contribute to the integral in \eqref{induced_int}, because $\hbar v/d$ corresponds to the X-ray energy scale. The mean-field Hamiltonians at different ${\bm k}$ therefore decouple from each other, leading to highly nonuniform gap function $\Delta_{\bm k}$ in the momentum space. We first consider the isotropic cavity, as defined in Sec.~\ref{2dcavity}. The result is shown in Fig.~\ref{fig:fs}(a) for $g_{\rm eff}^2=0.5$, where the superconducting gap parameter $\Delta_{\bm k}$ determined by the self-consistent calculation is plotted as a false color map. At large coupling strength, superconducting gaps emerge near the boundary of the FBZ. We then consider the fixed-polarization cavity. In this case, the BCS superconducting gap again emerges under strong coupling as shown in Fig.~\ref{fig:pair}(b). However, the gap now opens near the corners of the FBZ due to the strongly modified Fermi surface. In summary, the cavity-induced gap $\Delta_{\bm k}$ is highly momentum-dependent and only emerges at several discrete spots in the First Brillouin zone. The discrete gaps should then still be connected by segments of Fermi arcs. To demonstrate this scenario, we plot the density of states at zero energy, i.e., $A_{\bm k}(0)=-\frac{1}{\pi}\operatorname{Im}G^r_{\bm k}(0)$, in Fig.~\ref{fig:fs}. Physically, the strong momentum dependence indicates the real-space pairing $\Delta(\bm r_i-\bm r_j)=\sum_{\bm k} e^{i\bm k\cdot(\bm r_i-\bm r_j)}\Delta_{\bm k}/N$ can extend to spatial large distance $|\bm r_i-\bm r_j|$, which is consistent with the large force range of the cavity-mediated interaction. 

The paired state with segments of ungapped Fermi surface is generally less robust than the fully gapped local BCS pairing. Beyond the mean-field analysis, various fluctuations as well as the dimensional crossover can turn this state into a short-range ordered phase. In addition, this state should compete with the possible PDW state induced by the current-type interaction (the Amperean pairing). A complete analysis of these scenarios will be reserved for the future. 

\section{Photon undressing in a Fabry-P\'{e}rot cavity}
\label{sec:FP}

\begin{figure}
\includegraphics[scale=0.45]{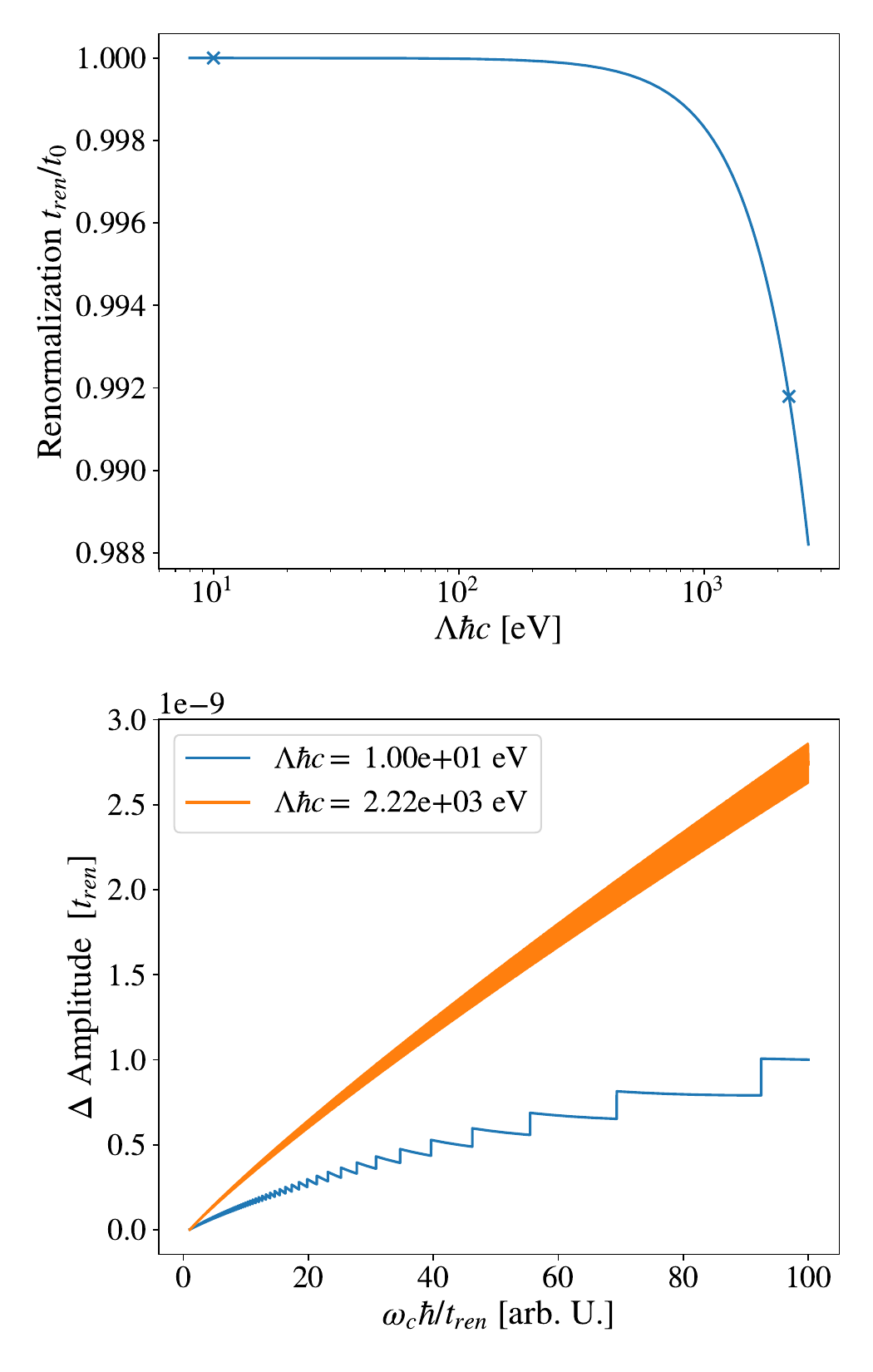}
\caption{Renormalization factor and induced hopping for $t_0=10$meV. The upper panel shows the dressed nearest-neighbor hopping in the unit of bare hopping $t_0$, depending on the ultraviolet cutoff $\Lambda$. The lower panel shows the change in the next-nearest hopping from the reference point $t_0=\omega_c$, measured in units of $t_{ren}(\omega_c=t_0)$. The kinks come from summation of modes below $\Lambda$. }
\label{fig:10meV}
\end{figure}

\begin{figure}
\includegraphics[scale=0.45]{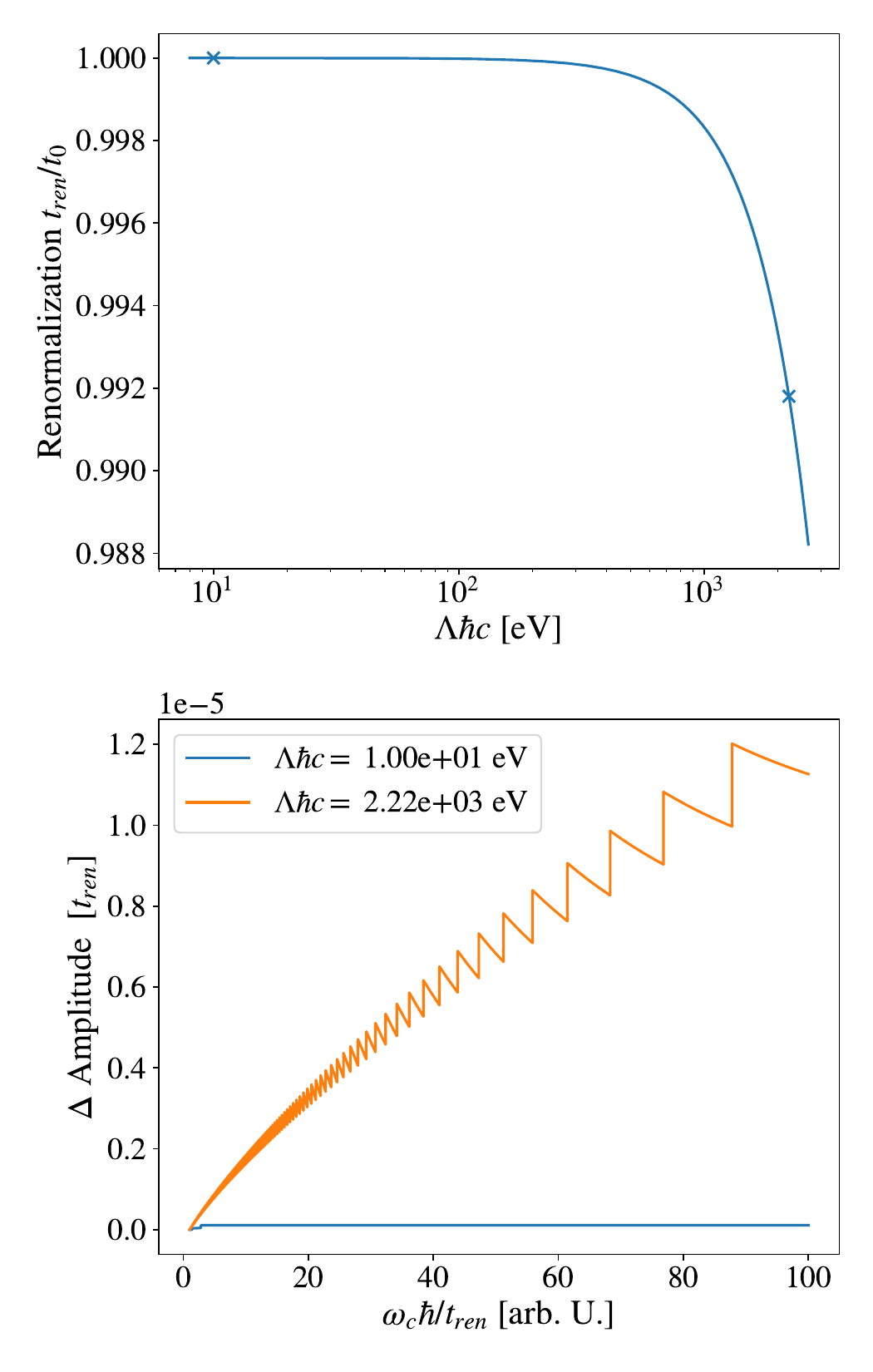}
\caption{The renormalization factor and induced hopping for $t_0=1$eV. Both panels show same results as in Fig.~\ref{fig:10meV} with the different bare hopping $t_0=1$eV. }
\label{fig:1000meV}
\end{figure}
The previous calculations have adopted a phenomenological method, where the effective coupling is freely varied to explore the hybrid light-matter phases. This setting corresponds to Situation I discussed in the introduction. In this section, we follow the {\it ab initio} strategy outlined as Situation II in the introduction, to estimate the light-matter coupling for a Fabry-P\'{e}rot cavity, which can be fabricated as a heterostructure by sandwiching the sample material between metallic layers. 

The Fabry-P\'{e}rot cavity selectively excludes modes whose wavelength $\lambda$ is not 
an integer multiple 
of $2L_z$, but does not strongly compress the retained modes like a subwavelength cavity. In the mathematical (though unphysical) limit $\omega_c\to \infty$, the cavity strips off all photon modes and the bare electron appears. Thus, both enhanced dressing of selected modes and undressing of unmatched modes should be treated on equal footing. This can lead to a distinct scenario from the bandwidth suppression in the previous sections. Although the effective coupling in the Fabry-P\'{e}rot cavity is very small ($\lesssim10^{-4}$), it is a nice example to show how a fully microscopic treatment can be carried out using the high-frequency formalism developed in this article. 

The smallness of the coupling justifies the neglect of higher-order contributions for each $n_z$ as well as all cross terms involving photon emission/absorption of modes with different $n_z$'s. The real Fabry-P\'{e}rot cavity features a rather complicated $\theta_{ij}(\bm q)$ factor in our formalism, which only contributes to a prefactor of magnitude $\sim 1$ in the results. In this section we aim to estimate the order of magnitude for the effects, and we will therefore assume constant $\theta$ functions in Eq.~\eqref{eq:pseudofp} for simplicity.

Specifically, we consider a square lattice with bare nearest-neighbor hopping $t_0$ placed between two infinite parallel mirrors and sum up the contribution from all modes with momenta $\bm q =(q_x, q_y, n_z\pi/L_z)$ with $|\bm q|<\Lambda$. As discussed before, we make the assumption that the effective model in the low-frequency limit defines the free-space dynamics, since the latter corresponds to a very large cavity ($\omega_c\to 0$). This limit cannot be reached in the high-frequency formulation, so we pick up $\omega_c=t_0$ as a reference point. The change of $t_{\rm NNN}$ from the reference point is measured and interpreted as the magnitude of the cavity-induced hopping. We evaluate the induced hopping up to the first-order term in $g_{\rm eff}^2$, in analogy to Eq.~\eqref{eq:hopping}, and sum up all contributions from modes with different $n_z$, taking $\omega_{n_z\bm q}=\sqrt{n_z^2\omega_c^2+v^2\bm q^2}$. \footnote{It is worth noting that the validity of this procedure is justified by the absence of infrared divergence in our calculation. Indeed, when $\omega_c\to 0$, the integral in \eqref{induced_int} leads to a factor of $\log(\Lambda/n_z q_c)/q_c$, multiplied with the factor $1/L_z^2\propto q_c^2$ in $g_{\rm eff}^2$. After summing over $n_z$, this gives rise to an infrared scaling of $\sum_{n_z}q_c\log(n_z q_c)\sim \int_{q_c} dq_z \log(q_z)$ as $q_c\to0$, which is convergent. It is worth noting that previous works have shown the infrared problem can be completely remedied in nonrelativistic quantum electrodynamics when the nonlinearity in the coupling term is retained \cite{hepp1973pra,rokaj2019}.}

In the following, we first calculate the renormalized nearest-neighbor hopping $t_{ren}$ and measure other quantities in the unit of  $t_{ren}(\omega=t_0)$ at the reference point. For practical calculations, we choose $t_0=10$meV and $1$eV, which are shown in Fig.~\ref{fig:10meV} and \ref{fig:1000meV}, respectively. The cutoff $\Lambda$ is varied between a typical plasma frequency for metals ($\Lambda \hbar c=10$eV) to the lattice momentum $\hbar c 2\pi/d=2.22\times10^3$eV, which corresponds to a lattice constant $d\approx 5.6$\r{A}. The dielectric constant is taken to be $\epsilon_r=13$, close to the value for GaAs \cite{schlawin2019}. For metallic mirrors, one should expect that the confinement only affects photons below their plasma frequency. 

The renormalized hopping $t_{ren}$, measured in the unit of bare $t_0$, exhibits a weak dependence on the cutoff $\Lambda$, leading to $r\sim 0.99$ up to lattice-momentum cutoff $\sim 2.22 \times 10^{3}$eV for the chosen parameters. The dependence of the renormalization on $\omega_c$ is significantly weaker, implying the band suppression effect is negligible in the conventional Fabry-P\'{e}rot cavity. However, if one can prepare a fixed-polarization cavity, the stripping of all modes in a fixed polarized direction still leads to an anisotropy, which may have a qualitative effect. The induced hopping varies between $10^{-9}$ to $10^{-5}$, which is relatively small,
 and consistent with the coupling strength estimated for individual dipoles placed in a Fabry-P\'{e}rot cavity \cite{devoret2007}. 
 
Finally, we note that the above analysis implies an interesting conclusion: The cavity-induced hopping increases with growing cavity frequency $\omega_c$. This can be understood by looking at the lowest-order term in Eq.~\eqref{eq:hopping}, which predicts an induced hopping $\propto\omega_c^2\log\Lambda^2$ for $\Lambda\gg \omega_c$. This observation is contradictory with the intuition from a single-mode approximation. In the latter case, the high-frequency contribution is proportional to $g^2/\omega_c\propto (1/\sqrt{\omega_c L_z})^2/\omega_c=1/\omega_c$, which decreases as $\omega_c$ grows. If one wants to simulate the continuum of modes with a single-mode toy model, the single-mode coupling has to grow as $g\propto\omega^{1.5}_c$ even for a conventional Fabry-P\'{e}rot cavity. To properly understand this effective enhancement of coupling, one has to take into account the continuum of modes. 

\section{Conclusion}
\label{sec:con}
In the article, we have formulated a high-frequency framework to study lattice electrons strongly coupled to quantum electromagnetic fields confined in a cavity, with an emphasis on cavity-induced control of electronic band structures and superconducting pairing. Our theory has mainly made progresses on two aspects: i) A continuum of cavity modes are taken into account, so that the thermodynamic limit can be properly reached, and the overall strength of the cavity-induced effects is not limited by the single-mode coupling \cite{rokaj2020}. ii) The nonlinear Peierls phase 
is treated nonperturbatively to maintain gauge-invariance, which is crucial for obtaining sensible conclusions in solid-state physics \cite{li2020prb}. With minimal assumptions on the cavity structure, we have derived a general effective model for the light-matter excitations and demonstrated a variety of intriguing physical consequences as listed below. 
\begin{enumerate}
\item The strong light-matter coupling induces electronic tunneling, modifying the band dispersion and Fermi surface. This effect offers the prospect of tuning competing orders, such as charge-density-wave and excitonic orders in quantum materials \cite{rettig2016}. Moreover, as a nonlinear effect due to the Peierls phase, the photon dressing also strongly suppresses the electronic hopping amplitudes along the polarized directions in the ultrastrong light-matter coupling regime. 
\item The cavity setup can break crucial symmetries in the system, such as the point-group symmetry (the fixed-polarization cavity) and the time-reversal symmetry (the chiral cavity). In particular, we have demonstrated that, This leads to a unique possibility to break the crystal symmetry and enhance the anisotropy in selected directions, leading to dimensional crossover in the ultrastrong coupling regime. Moreover, breaking of time-reversal symmetry can switch the topological class of certain material samples and induce topologically nontrivial states without applying strong external fields. 
\item The photon-mediated interaction gives rise to nonlocal paired states and segments of superconducting gaps in the momentum space. In particular, we show that, in the presence of nonlinear Peierls phase coupling, two types of interactions arise due to exchange of virtual photons. In addition to the current-type interaction which couples current fluctuations at different spatial positions, another kinetic-type interaction emerges due to the higher-order coupling terms and is absent under a linear truncation of the Peierls phase. It leads to a nonlocal but uniformly paired state (BCS pairing of electrons with opposite momenta) distinct from the PDW phases induced by current-type interactions. 
\end{enumerate}

Our study is based on a Brillouin-Wigner high-frequency expansion \cite{mikami2016}, which is most relevant to scenarios where the cavity frequency $\omega_c$ is much greater than the electronic energy scales, such as the electronic bandwidth. In solids, strict fulfillment of this assumption requires a fine tuning between the electronic dispersion and the cavity structure to avoid strong hybridization with other high-lying excitations and energy bands. However, the high-frequency approximation can still give reliable predictions to low-energy physics, e.g., near the Fermi surface, even if the cavity frequency is not significantly beyond the electronic bandwidth. 

We have also explored the cavity-induced effects in a Fabry-P\'{e}rot-type cavity consisting of two parallel mirrors. The cavity confinement leads to similar effects, such as induced hopping as in the ultrastrong coupling regime, while the strength of effects is relatively small. The effects can be enhanced for, e.g., subwavelength cavities and metamaterials or heterostructures \cite{kennes2021,topp2021}. Cold-atom systems represent another family of possible implementations \cite{yang2020,mil2020}, which can also be studied using the same formulation discussed in Sec.~\ref{sec:method}. A scalable implementation of subwavelength cavities or artificial dynamical gauge fields is therefore pivotal for implementing the cavity-induced band engineering.

Many directions can be explored using the formalism developed in this article. For example, fluctuations beyond the mean-field analysis carried out here can possibly enhance the cavity-induced effects even for a conventional Fabry-P\'{e}rot cavity, like observed in Dicke model. Moreover, we have concentrated on the hybridization of ideal photonic and electronic excitations. In particular, the cavity is assumed to be completely lossless, and the other degrees of freedom in solids, such as phonons, are ignored. However, dissipation plays 
a crucial role 
in both solids and realistic cavities, and electron-phonon coupling is especially important in quantum materials close to a phase transition \cite{kennes2017,grandi2020}. The questions can be phenomenologically addressed by coupling the effective models to external baths and solving them using Green's function methods, similar to the Floquet Green's function formalism \cite{tsuji2008,kitagawa2011}. In addition, nonequilibrium protocols can be a useful tuning knob for the cavity-matter systems, which have not been considered in this article. The interplay between the cavity-mediated interaction and the static Coulomb interaction is also not discussed here, but is a very interesting question that should be examined in the future. Finally, the low-frequency regime (with $\omega_c\lesssim t_0$) can be addressed in a field-theoretic or diagrammatic formalism, while the high-frequency theory can still provide a benchmark for specific cases.

\begin{acknowledgments}
We thank G.~Mazza, X.~Wang, A.~J.~Kim, and P.~Werner for inspiring discussions. J.~L. and M.~E. were funded by the ERC Starting Grant No.~716648,
and  by the Deutsche Forschungsgemeinschaft (DFG, German Research Foundation) -- Project-ID 429529648 -- TRR 306 QuCoLiMa ("Quantum Cooperativity of Light and Matter''). J.~L. thanks the funding from the European Union’s Horizon 2020 research and innovation programme under the Marie Sk\l{}odowska-Curie grant agreement No. 884104.
\end{acknowledgments}
\appendix

\section{Evaluation of the Brillouin-Wigner series}

We follow the convention in Ref.~\citenum{li2020prl} and define the polynomial $j_{nm}$ by evaluating the matrix element of the Peierls phase 
\begin{align}
\label{jnm}
j_{n,m}(2x)&=\bra{n}e^{ix(a+a^\dag)}\ket{m}\nonumber\\
&=e^{-x^2/2}\sum_{k=0}^m\frac{(-1)^k x^{2k+|n-m|}}{k!(k+|n-m|)!}\sqrt{\frac{n!}{m!}}\frac{m!}{(m-k)!},
\end{align}
with $j_{nm}=j_{mn}$.

The Brillouin-Wigner perturbation series can be carried out order by order. The key observation is that, for a continuum of intermediate states labelled by $\bm q$, the expressions are highly simplified since many terms become zero in the thermodynamic limit $L_x\sim L_y\to\infty$. For simplicity of notation, we define $S=L_xL_y$. 

\subsection{Zeroth-order term}
As in the main text, we first compute the renormalization factor as an example,
\onecolumngrid
\begin{align}
\prod_{\bm  q}j_{n_{\bm  q},n_{\bm  q}}(2|g_{ij}(\bm  q)|)&=e^{-(2\pi)^2\sum_{\bm  q} |\gamma_{Ij}({\bm  q})|^2/2L_c^2}\left(1-(2\pi)^2\sum_{\bm  q}n_{\bm  q} \frac{|\gamma_{ij}(\bm  q)|^2}{S}+(2\pi)^4\frac{1}{2}\sum_{\bm  q_1\ne\bm q_2}n_{\bm  q_1}n_{\bm  q_2} \frac{|\gamma_{ij}(\bm  q_1)|^2|\gamma_{ij}(\bm  q_2)|^2}{S^2}+\ldots \right)\nonumber\\
&=e^{-\int^\Lambda d^2\bm q |\gamma_{ij}(\bm  q)|^2/2}\left(1-\int^\Lambda d^2\bm  q n_{\bm  q} |\gamma_{ij}(\bm  q)|^2+\frac{1}{2}\iint^\Lambda d^2\bm  q_1
d^2
\bm q_2 n_{\bm  q_1} n_{\bm  q_2} |\gamma_{ij}(\bm  q_1)|^2|\gamma_{ij}(\bm  q_2)|^2+\ldots \right)\nonumber\\
&=\exp[{-\int^\Lambda d^2\bm q \left(n_{\bm q}+\frac{1}{2}\right)|\gamma_{ij}(\bm  q)|^2}],
\end{align}
\twocolumngrid
where the general $k$th term is 
\begin{align}
\frac{(-1)^k}{k!}\prod_{s=1}^k z_{ij}^2\int d \bm q_s n_{\bm q_s} |\gamma(\bm  q_s)|^2.
\end{align}
As stated in the main text, the $\lambda$ subscript has been omitted. All integrals are assumed to be limited by $\Lambda$ in the following. Note that for the first equality all terms of order $\mathcal{O}(1/S)$ are omitted. In particular, for the \emph{low-energy} sector ($n_{\bm  q}=0$) we have 
\begin{align}
\prod_{\bm  q}j_{0,0}(2|g_{ij}(\bm  q)|)= \exp({-\int d^2\bm q  \frac{1}{2}|\gamma_{ij}(\bm  q)|^2}).
\label{geff}
\end{align}

\subsection{Higher-order terms}

To calculate the first order term, we note $j_{n,m}(g)\sim g^{|n-m|}$,
so that $j_{n,n+l} j_{n+l,n}\sim 1/S^{l}$, and a term involving this factor will need $l$ summations ($\sum_{\bm q}$). We can slightly rewrite the expansion as follows, 
\begin{align}
\sum_{ \bm l\ne \bm 0}\frac{\mathcal{H}_{\bm n, \bm n+\bm l}\mathcal{H}_{\bm n+\bm l,\bm n}}{\bm l \cdot \bm  \omega}.
\end{align}
This expression can be ill-defined for $\bm n \ne \bm 0$, since two \emph{degenerate} photon modes can enter the same denominator, leading to $ \omega_{\bm  q_1}- \omega_{\bm  q_2}=0$. Even if the photon modes are nondegenerate, in the continuum-mode limit one has arbitrary small $ \omega_{\bm  q_1}- \omega_{\bm  q_2}$ and the expansion breaks down. However, in two situations this divergence can be avoided:
\begin{itemize}
\item In the low-energy sector ($\bm n = \bm 0$). Any other sectors are separated from the low-energy sector by at least a photon excitation $ \omega_c$.
\item In the strong electronic correlation limit. For example, consider a Hubbard lattice with $U\gg t_0$ and assume \emph{non-resonance} condition $U\ne n_{\bm q} \omega_{\bm q}$ for any mode $\bm q$ and $n_{\bm  q}\in \mathbb{Z}^+$. In this case, the high frequency limit is reinterpreted as $ \omega_c\gg J_{\rm ex}$, and the non-resonance condition must be satisfied.
\end{itemize}
Indeed, the first case is protected by the excitation gap $ \omega_c$, and the second case is protected by the interaction gap (charge gap for $U>0$ and spin gap for $U<0$ in the Hubbard model). In the general situation, nevertheless, excited electronic states and photon states with similar energies can strongly hybridize with each other and 
using the high frequency expansion to derive 
an effective theory for a given photon-number sector labeled by $\bm n$ is ill-defined. We stress that, in the free-space limit where $ \omega_c\to 0$, 
this holds even for the sector $\bm n=0$.

To consider the low-energy effective model, we fix $\bm n =\bm 0$ and compute $\sum_{ \bm l > \bm 0}\frac{\mathcal{H}_{\bm 0, \bm l}\mathcal{H}_{\bm l,\bm 0}}{\bm l \cdot \bm  \omega}$. We note that, for a fixed label $\bm l$, $ \mathcal{H}_{\bm 0, \bm l}\mathcal{H}_{\bm l,\bm 0}\sim \mathcal{O}(1/S^{|\bm l|})$, where $|\bm l|=\sum_{\bm  q}l_{\bm q}$. To reach a nonzero limit, the term $\mathcal{H}_{\bm 0, \bm l}\mathcal{H}_{\bm l,\bm 0}$ should be put inside $|\bm l|$ mode-summations (which introduce a factor $S^{|\bm l|}$).
The expansion can be organized w.r.t to $|\bm l|$. For a fixed $|\bm l|$, each entry of $\bm l$ must satisfy $l_i\le1$ for $i=1, \ldots, M$ so that there are precisely $|\bm l|$ summations. We define $\bm l(\bm q_1,\ldots, \bm q_N)$ to be a label whose $l_{\bm q_s}=1$ for $s=1,\ldots,N$ and other components vanish and the term then becomes
\begin{align}
\sum_{|\bm l| \ge 1}\frac{\mathcal{H}_{\bm 0, \bm l}\mathcal{H}_{\bm l,\bm 0}}{\bm l \cdot \bm  \omega}&=\sum_{\bm q}\frac{\mathcal{H}_{\bm 0, \bm l(\bm  q)}\mathcal{H}_{\bm l(\bm  q),\bm 0}}{ \bm  \omega_{\bm q}}+\nonumber\\
&\quad\frac{1}{2}\sum_{\bm q_1\ne\bm q_2}\frac{\mathcal{H}_{\bm 0, \bm l(\bm  q_1,\bm q_2)}\mathcal{H}_{\bm l(\bm  q_1,\bm q_2),\bm 0}}{ \bm  \omega_{\bm q_1}+\bm  \omega_{\bm q_2}}+\ldots.
\end{align}
The general term of order $|\bm l|$ is, therefore, $\frac{1}{|\bm l|!}\sum_{\{\bm  q_s\}}{\mathcal{H}_{\bm 0, \bm l(\{\bm q_s\})}\mathcal{H}_{\bm l(\{\bm q_s\}),\bm 0}}/{\sum_s \omega_{\bm  q_s}}\sim\mathcal{O}(1)$. This term will involve products like below (for example, when $|\bm l|=1$)
\begin{align}
j_{00}(|g(1)|)j_{00}(|g(2)|)\times\ldots \times j_{01}(|g(\bm q_s)|)\times \ldots \times j_{00}(|g(M)|),
\end{align}
where $g(a)$ represents the coupling to the mode $a$ and the $j_{nm}$ function is given by \eqref{jnm}.
To evaluate the general term, recall that $j_{n,n+1}(2|g_{ij}|) = e^{-|g_{ij}|^2/2} \sqrt{n+1} |g_{ij}|(1-\frac{1}{2}n |g_{ij}|^2+\mathcal{O}(|g|^4))$. For the low-energy sector, we have $n=0$ so $j_{0,0}=e^{-|g_{ij}|^2/2}(1+\mathcal{O}(|g|^4))$ and $j_{0,1}=e^{-|g_{ij}|^2/2}|g_{ij}|(1+\mathcal{O}(|g|^4))$. Using these expressions one can obtain the effective Hamiltonian \eqref{induced_int}.

This gives the full result of the induced interaction. An interesting observation is that the $|\bm l|$th term decays roughly as $1/|\bm l|!$, indicating the shrinking phase-space volume for higher-order scattering. The power-counting argument can be understood with a simple physical interpretation: for the single-mode coupling, a virtual emission/absorption can involve an arbitrary number of photons from the \emph{same} mode. Although this is impossible for the continuum-mode limit since $g\propto 1/\sqrt{S}$, it is possible to absorb/emit many virtual photons from \emph{different} modes.

Finally, we note that one can also consider more than one branches of photons, such as different polarizations and higher longitudinal modes with different $\omega_c$'s, labelled by $\lambda$ in the original Hamiltonian \eqref{ham}. In this case, one has to restore the other indices and sum over different modes on top of the $\bm q$ summation, leading to general term as follows,

\begin{align}
\frac{1}{|\bm l|!}\sum_{\{\bm  q_s\lambda_s\}}{\mathcal{H}_{\bm 0, \bm l(\{\bm q_s\lambda_s\})}\mathcal{H}_{\bm l(\{\bm q_s\lambda_s\}),\bm 0}}/({\sum_s \omega_{\bm  q_s\lambda_s}}),
\label{morebranch}
\end{align}
where $\lambda_s$ can represent a general quantum number such as $q_z$ and polarization.  

\section{The induced hopping}
\label{appB}
In this section, we compute the induced hopping due to the cavity confinement. We first consider a single branch of photon modes (one set of $\bm q$ modes) and then generalize it to the case of two polarizations with degenerate frequencies ($\omega_{\bm q,1}=\omega_{\bm q,2}=\omega_{\bm q}$). The induced hopping is given by $V^{ij}_{jk}$ in Eq.~\eqref{induced_int}. We note that, due to the large value of speed of light, the cavity momentum $q_c=\omega_c/v$ is generally much smaller than the lattice momentum cutoff $\sim 2\pi/d$, where $d$ represents the magnitude of the lattice constant. As a result, the $\exp(i \bm q\cdot \Delta\bm R)$ factor in $V^{ij}_{jk}$ can be ignored. Since we do not want to model the field strength in a cavity from its microscopic description, we will adopt a similar expression of $\gamma_{ij}$ given as Eq.~\eqref{eq:gamma} and assume the coupling constant is enhanced to the ultrastrong coupling regime by an additional subwavelength compression factor $A$. The result reads,
\begin{align}
\frac{\omega_c V^{ij}_{jk}}{r_{ij}r_{jk}}&\approx\omega_c\sum_{l=1}^{\infty}\frac{(-)^{l}d^{2l}}{l!} \int d^2\{\bm q_{ s}\} \frac{\prod_{s=1}^{l} \gamma^*_{ij}(\bm q_{ s})\gamma_{jk}(\bm q_{ s})}{\sum_{s=1}^{l} \omega_{\bm q_{ s}}}\nonumber\\
&=\sum_{l=1}^{\infty}\frac{(-)^{l}(2A\alpha d^2)^l}{(\lambda_c^2 \sqrt{\epsilon_r})^ll!}\int \frac{d^2\{\bm q_{ s}\}}{q_c^{2l}} \prod_{s=1}^l\frac{\theta^*_{ij}(\bm q_{ s})\theta_{jk}(\bm q_{ s})}{\tilde{\omega}_{\bm q_{ s}} \sum_{s=1}^l\tilde{\omega}_{\bm q_{ s}}}\nonumber\\
&=\sum_{l=1}^{\infty}\frac{(-)^{l}g^{2l}_{\rm eff}}{l!}\int \frac{d^2\{\bm q_{ s}\}}{q_c^{2l}} \prod_{s=1}^l\frac{\theta^*_{ij}(\bm q_{ s})\theta_{jk}(\bm q_{ s})}{\tilde{\omega}_{\bm q_{ s}} \sum_{s=1}^l\tilde{\omega}_{\bm q_{ s}}}/\Gamma^l, 
\end{align}
where the dimensionless $\tilde{\omega}_{\bm q}=\sqrt{1+|\bm q|^2/q_c^2}$ and $\Gamma=\sqrt{1+\Lambda^2/q_c^2}-1$, and the effective coupling $g_{\rm eff}$ given by Eq.~\eqref{eq:geff}). When $\theta_{ij}$ only depends on the direction of $\bm q$, one can factor out the angular integral of $\bm q$ and define $z_{ij, jk}=\int d\varphi \theta^*_{ij}(\varphi)\theta_{jk}(\varphi)/2\pi$. The integration can then be carried out explicitly, 

\begin{align}
&\sum_{l=1}^{\infty}(2\pi)^l \frac{g^{2l}_{\rm eff} z^l_{ij,jk}}{l!}\times\nonumber\\
&\qquad \int_0^{\Lambda/q_c} \prod_{s=1}^l\frac{ 2p_s dp_s}{\sqrt{1+p_s^2}}\frac{1}{ \sum_{s=1}^l\sqrt{1+p_s^2}}/\Gamma^l,\nonumber\\
&=f(- z_{ij,jk}g^2_{\rm eff}),
\label{ffunc}
\end{align} 
where $p_s=|\bm q|/q_c$. In the main text, we indeed treat $g_{\rm eff}$ as a free parameter. The integral on the second last line roughly increases as $\left(\sqrt{1+\Lambda^2/q_c^2}-1\right)^{l-1}\ln\sqrt{1+\Lambda^2/q_c^2}$ with UV cutoff $\Lambda$. An analytic expression can be obtained for $f(x)=\sum_{l=1}^\infty C_l(2\pi x)^l/l!$ where the coefficients are given as follows,
\begin{align}
C_l&=\sum_{i=0}^l {l\choose i}\frac{(-1)^i}{(l-1)!}\left((l-i)+i/\Gamma\right)^{l-1}\times\nonumber\\
&\quad\left(\ln((l-i)(1+\Gamma)+i)-\sum_{k=1}^{l-1}\frac{1}{k}\right)/\Gamma.
\end{align}
As stated in the beginning of the section, this expression is valid for 2D systems coupled to a single branch of $\bm q$ modes. Usually we choose the $z$--momentum $q_z=q_c$ ($n_z=1$) since it contributes to the effective Hamiltonian most strongly in the high-frequency regime. Other branches with different $q_z$ and polarizations can be included using Eq.~\ref{morebranch}. For example, when two polarizations $\lambda=1,2$ are included for each momentum, the integration on the first line of \eqref{ffunc} must be equipped with an extra summation for $\lambda$. In particular, when frequencies are degenerate for different polarizations, the denominator in \eqref{morebranch} becomes identical for different combination of polarization $\lambda$. In the situation discussed in the main text, only the $\theta$ factor depends on the polarization, and the $l$th-order term can then be rearranged as follows, 

\begin{align}
&\sum_{\lambda_s}\int d^2\{\bm q_{ s}\} \frac{\prod_{s=1}^{l}\gamma^*(q_{ s})\gamma(q_{ s})\theta^*_{ij}(\varphi_{ s},\lambda_{ s})\theta_{jk}(\varphi_{ s},\lambda_{ s})}{\sum_{s=1}^{l} \omega_{\bm q_{ s}\lambda_s}}\nonumber\\
&=(2\pi)^l\int \prod_s q_s dq_s \frac{\prod_{s=1}^{l}\gamma^*(q_{ s})\gamma(q_{ s})}{\sum_{s=1}^{l} \omega_{\bm q_{ s}}}\times\nonumber\\
&\qquad(\sum_{\lambda=1}^{2} \int \frac{d\varphi}{2\pi} \theta^*_{ij}(\varphi,\lambda)\theta_{jk}(\varphi,\lambda))^l.\nonumber\\
&=(2\pi)^l (\sum_{\lambda}g^{2}_{\rm eff,\lambda} z_{ij,jk;\lambda})^l\times\nonumber\\
&\qquad \int_0^{\Lambda/q_c} \prod_{s=1}^l\frac{ 2p_s dp_s}{\sqrt{1+p_s^2}}\frac{1}{ \sum_{s=1}^l\sqrt{1+p_s^2}}/\Gamma^l,\nonumber\\
\end{align}
This leads to an induced hopping of the following form $f(- \sum_{\lambda}z_{ij,jk;\lambda}g^2_{\rm eff,\lambda})$. In other words, different polarizations (and any other quantum numbers with degenerate frequencies) should be summed up as the argument of the $f$ function. This is also expected on physical grounds, because the effective Hamiltonian should be invariant under linear basis changes of degenerate polarization states.
\bibliography{cavity.bib}

\end{document}